\newcommand*{\ATLASLATEXPATH}{}
\pdfoutput=1
\documentclass[cernpreprint,txfonts,USenglish,texlive=2016]{\ATLASLATEXPATH atlasdoc}
\usepackage[subfigure]{\ATLASLATEXPATH atlaspackage}
\usepackage{\ATLASLATEXPATH atlasbiblatex}
\usepackage{\ATLASLATEXPATH atlascontribute}
\usepackage{\ATLASLATEXPATH atlasphysics}

\usepackage{paper_BLstop2016-defs}
\usepackage{rotating}

\AtlasTitle{A search for $B-L$ $R$-parity-violating top squarks in $\sqrt{s} = 13$~$\text{Te\kern -0.1em V}$ $pp$ collisions with the ATLAS experiment}
\author{The ATLAS Collaboration}
\AtlasRefCode{SUSY-2016-29}
\PreprintIdNumber{CERN-EP-2017-171}
\AtlasJournalRef{Phys. Rev. D 97 (2018) 032003}
\AtlasDOI{10.1103/PhysRevD.97.032003}
\AtlasAbstract{%
  A search is presented for the direct pair production of the stop, the supersymmetric partner of the top quark, that decays through an
  $R$-parity-violating coupling to a final state with two leptons and two jets, at least one of which
  is identified as a $b$-jet.
  The dataset corresponds to an integrated luminosity of 36.1~fb$^{-1}$ of proton--proton collisions at a
  center-of-mass energy of $\sqrt{s} = 13$~$\text{Te\kern -0.1em V}$, collected in 2015 and 2016 by the ATLAS detector at the LHC.
  No significant excess is observed over the Standard Model background, and exclusion limits are set on
  stop pair production at a 95\% confidence level.
  Lower limits on the stop mass are set between 600~$\text{Ge\kern -0.1em V}$ and 1.5~$\text{Te\kern -0.1em V}$ for branching ratios above 10\% for
  decays to an electron or muon and a $b$-quark.
}

\hypersetup{pdftitle={ATLAS document},pdfauthor={The ATLAS Collaboration}}

\begin{document}

\newcommand{\Ltab}[1]{\multicolumn{1}{l}{{ }{ }#1}}
\newcommand{\LtabOne}[1]{\multicolumn{1}{l}{#1}}

\maketitle

\tableofcontents
\clearpage

\section{Introduction}
\label{sec:intro}

The extension of the Standard Model (SM) of particle physics with
supersymmetry
(SUSY)~\cite{Golfand:1971iw,Volkov:1973ix,Wess:1974tw,Wess:1974jb,Ferrara:1974pu,Salam:1974ig}
leads to processes that violate both baryon number ($B$)
and lepton number ($L$), such as rapid proton decay.
A common theoretical approach to reconcile the strong constraints from the non-observation of these
processes is to introduce a multiplicative quantum number called
$R$-parity~\cite{Farrar:1978xj},
defined as $R=(-1)^{3(B-L)+2s}$ where $s$ is the spin of the
particle. If $R$-parity is conserved, then SUSY particles are produced
in pairs, and the lightest supersymmetric particle (LSP) is stable.
The LSP cannot carry electric charge or color charge without coming
into conflict with astrophysical data~\cite{Ellis:1983ew,Pospelov:2006sc}.

A number of theoretical models beyond the Standard Model (BSM) predict $R$-parity violation (RPV)~\cite{Dreiner:1997uz,Barbier:2004ez,Perez:2013kla,Restrepo:2001me}.
The benchmark model for this search considers an additional local symmetry
$U(1)_{\BL}$ to the $SU(3)_\mathrm{C} \times SU(2)_\mathrm{L} \times U(1)_\mathrm{Y}$ Standard
Model with right-handed neutrino supermultiplets. The minimal supersymmetric
extension then only needs a vacuum expectation value for a
right-handed scalar neutrino in order to spontaneously break the \BL
symmetry~\cite{FileviezPerez:2008sx, Barger:2008wn, Everett:2009vy,
Braun:2005ux, Deen:2016vyh}.  This
minimal \BL model violates lepton number but not baryon number. The
couplings for RPV are highly suppressed as
they are related to the neutrino masses, and the model is consistent
with the experimental bounds on proton decay and lepton number
violation.  At the LHC, the most noticeable effect is that the LSP is
no longer stable and can now decay via RPV processes,
and it also may now carry color and electric charge.  This leads to
unique signatures that are forbidden in
conventional models with $R$-parity conservation.
A novel possibility is a top squark or \Stop (\StopSym) as the LSP with a rapid RPV decay.
The supersymmetric partners of the left- and right-handed top quarks, $\StopSym_{\mathrm{L}}$ and $\StopSym_{\mathrm{R}}$,
mix to form two mass eigenstates consisting of the lighter $\StopSym_{1}$ and heavier $\StopSym_{2}$.
Given the large top quark mass, the lighter $\StopSym_{1}$ is expected to be
significantly lighter than the other squarks due to renormalization
group effects~\cite{Barbieri:1987fn, deCarlos:1993rbr}.
The lighter $\StopSym_{1}$, denoted \StopSym for simplicity, is the target of this analysis.

\begin{figure}[t]
  \centering{
    \includegraphics[scale=0.75]{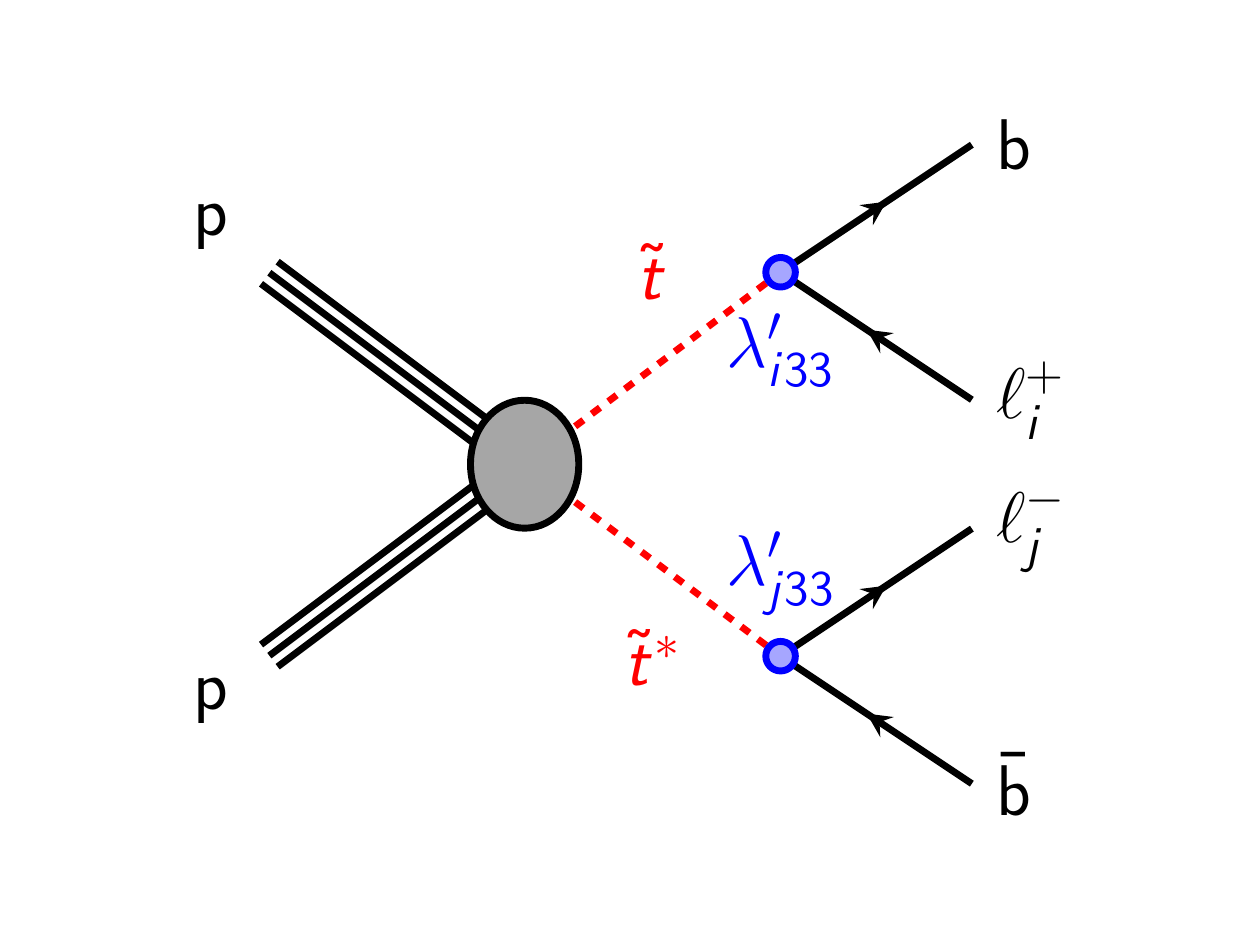}
  }
  \caption{Feynman diagram for \Stop pair production, with \StopSym and anti-\StopSym ($\StopSym^{*}$)
    decay to a charged lepton of any flavor and a $b$-quark through an $R$-parity-violating coupling $\lambda^{\prime}$.}
  \label{fig:feynman_diagram}
\end{figure}

This paper presents a search performed by ATLAS for
direct \Stop pair production, with the RPV decay of each \StopSym to a
$b$-quark and a charged lepton ($\StopSym~\to~\bl$), as shown in
\figref{feynman_diagram}.  In contrast to $R$-parity-conserving searches
for \StopSym, there is no significant missing
transverse momentum in the decay. The \StopSym decay branching ratios to each
lepton flavor are related to the neutrino mass hierarchy~\cite{Marshall:2014cwa, Marshall:2014kea},
and a large phase space in the branching ratio plane is currently available.
With an inverted mass hierarchy the branching ratio to the $be$ final state may be as
large as 100\%, and with a normal mass hierarchy the branching ratio
to the $b\mu$ final state may be as high as 90\%. The experimental signature
is therefore two oppositely charged leptons of any flavor and two $b$-jets.
In this analysis, only events with electron or muon signatures are
selected, and final states are split by flavor into \EE, \EM, and \MM selections.
At least one of the two jets is required to be identified as initiated by a $b$-quark, improving
the selection efficiency of signal events over a requirement of two $b$-jets.
Events are chosen in which the two reconstructed \bl pairs have roughly equal mass.

Previous searches with similar final states have targeted
the pair production of first-, second-, and third-generation leptoquarks
at ATLAS~\cite{EXOT-2015-19,EXOT-2014-03}
and at CMS~\cite{CMS-EXO-16-023,CMS-EXO-12-041}.  However, they consider final
states within the same generation ($ee jj$, $\mu \mu jj$, $\tau \tau
bb$, where $j$ indicates a light-flavor jet) and do not focus on final states
with both $b$-jets and electrons or muons ($ee bb$, $\mu \mu bb$), nor
consider final states with both electrons and muons ($e \mu bb$).  The
results of the Run 1 leptoquark searches were reinterpreted for
the \StopSym mass and its decay branching ratios in the \BL
model~\cite{Marshall:2014cwa, Marshall:2014kea}, setting lower mass limits
between 424 and 900~\GeV at a 95\% confidence level.

The ATLAS detector and the dataset collected during Run 2 of the LHC are described in \secref{detector},
with the corresponding Monte Carlo simulation samples presented in \secref{mc}.
The identification and reconstruction of jets and leptons is presented in \secref{reco},
and the discriminating variables used to construct the signal regions are described in \secref{sel}.
The method of background estimation is described in \secref{bkg}, and the systematic uncertainties are detailed in \secref{systematics}.
The results are presented in \secref{result}, and the conclusion given in \secref{conclusion}.

\section{ATLAS detector and data set}
\label{sec:detector}
The ATLAS detector~\cite{PERF-2007-01} consists of an inner detector tracking system,
electromagnetic and hadronic sampling calorimeters, and a muon spectrometer.
Charged-particle tracks are reconstructed in the inner detector (ID), which spans the
pseudorapidity\footnote{ATLAS uses a right-handed coordinate system with its origin at the nominal interaction point (IP) in the center of the detector and the $z$-axis along the beam pipe. The $x$-axis points from the IP to the center of the LHC ring, and the $y$-axis points upward. Cylindrical coordinates $(r,\phi)$ are used in the transverse plane, $\phi$ being the azimuthal angle around the $z$-axis. The pseudorapidity is defined in terms of the polar angle $\theta$ as $\eta=-\ln\tan(\theta/2)$.  Rapidity is defined as $y=0.5\ln\left[\left(E+p_{z}\right)/\left(E-p_{z}\right)\right]$ where $E$ denotes the energy and $p_{z}$ is the component of the momentum along the beam direction.} range $|\eta|<2.5$, and consists of three subdetectors:
a silicon pixel tracker, a silicon microstrip tracker, and a straw-tube transition radiation tracker.
The ID is surrounded by a thin superconducting solenoid providing an axial magnetic field of 2~T, allowing the measurement of charged-particle momenta.
In preparation for Run~2, a new innermost layer of the silicon pixel tracker, the insertable B-layer (IBL)~\cite{ATLAS-TDR-19},
was introduced at a radial distance of 3.3~cm from the beamline to improve track reconstruction
and the identification of jets initiated by $b$-quarks.

The ATLAS calorimeter system consists of high-granularity electromagnetic and hadronic sampling calorimeters covering the region $|\eta|$ < 4.9.
The electromagnetic calorimeter uses liquid argon (LAr) as the active material with
lead absorbers in the region $|\eta|$ < 3.2.
The central hadronic calorimeter incorporates plastic scintillator tiles and steel absorbers in the region $|\eta|$ < 1.7.
The hadronic endcap calorimeter (1.5 < $|\eta|$ < 3.2) and the forward calorimeters (3.1 < $|\eta|$ < 4.9) use LAr with copper or tungsten absorbers.

The muon spectrometer (MS) surrounds the calorimeters and measures muon tracks within $|\eta|$ < 2.7
using three layers of precision tracking chambers and dedicated trigger chambers.
A system of three superconducting air-core toroidal magnets provides a magnetic field for measuring muon momenta.

The ATLAS trigger system begins with a hardware-based level-1 (L1) trigger followed by a
software-based high-level trigger (HLT)~\cite{TRIG-2016-01}.
The L1 trigger is designed to accept events at an average rate of 100~kHz, and the HLT is designed to accept events to write out to disk at an average rate of 1~kHz.
Electrons are triggered in the pseudorapidity range $|\eta|$ < 2.5, where the electromagnetic calorimeter is finely segmented and track reconstruction is available.
Compact electromagnetic energy deposits triggered at L1 are used as the seeds for HLT algorithms that
are designed to identify electrons based on calorimeter and fast track reconstruction.
The muon trigger at L1 is based on a coincidence of trigger chamber layers.
The parameters of muon candidate tracks are then derived in the HLT by fast reconstruction algorithms in both the ID and MS.

The data sample used for this search was collected from proton--proton collisions at a
center-of-mass energy of $\sqrt{s}=13~\TeV$ in 2015 and 2016.
An integrated luminosity of 36.1~\ifb\ was collected while all tracking detectors, calorimeters, muon chambers, and magnets
were fully operational.
The uncertainty in the combined 2015 and 2016 integrated luminosity is 3.2\%.
It is derived from a preliminary calibration of the luminosity scale using $x$--$y$ beam-separation scans performed in August 2015 and May 2016,
following a methodology similar to that detailed in Ref.~\cite{DAPR-2013-01}.
The LHC collided protons with bunch-crossing intervals of 25~ns, and the
average number of interactions per bunch crossing was estimated to be $\avg{\mu}=23.7$.

For this analysis, events are selected using single-electron and single-muon triggers requiring leptons above a transverse momentum (\pt) threshold and satisfying various lepton identification and isolation criteria.
The trigger-level criteria for the \pt, identification, and isolation of the leptons are less stringent
than those applied in the event selection to ensure that trigger efficiencies are constant in the analysis phase space.

\section{Monte Carlo simulation}
\label{sec:mc}
Monte Carlo (MC) simulation is used to predict the backgrounds from SM processes,
estimate the detector response and efficiency to reconstruct the signal process, and estimate systematic uncertainties.
The largest sources of SM background with different-flavor leptons are top quark pair production (\ttbar) and single-top-quark production (\singletop),
while the largest source with same-flavor leptons is \Zjet production,
The yields of these three backgrounds are estimated through data-driven methods described in \secref{bkg}.
The smaller backgrounds are \Wjet, diboson, and \ttbar+$W/Z$ production and are estimated directly from MC simulation.
The contribution from events with jets misreconstructed as leptons or with non-prompt leptons is evaluated with the MC simulation and is negligible.
Details of the MC simulations are given below and are summarized in \tabref{MC}.

\begin{table}[t]
\footnotesize
\centering{
\caption{MC simulation details by physics process.}
\label{tab:MC}
  \begin{tabular}{lccccc}
  \toprule
    Process & Event & PS and & UE tune & PDF & Cross section \\
    & generator & hadronization & & & \\
    \hline
    \hline

    \ttbar   & \POWHEGBOX~v2 & \PYTHIAV{6.428} & P2012 & CT10  & NNLO+NNLL~\cite{Czakon:2011xx} \\
    \singletop & & & & \\
    (\Wt and $s$-channel) & \POWHEGBOX~v2 & \PYTHIAV{6.428} & P2012 & CT10 & NNLO+NNLL~\cite{Kidonakis:2010ux,Kidonakis:2010tc} \\
    ($t$-channel) & \POWHEGBOX~v1 & \PYTHIAV{6.428} & P2012 & CT10 & NNLO+NNLL~\cite{Kidonakis:2011wy} \\
    $Z/W$+jets & \Sherpa 2.2.1 & \Sherpa 2.2.1 & Default & NNPDF3.0 & NNLO~\cite{Catani:2009sm} \\
    Diboson  & \Sherpa 2.2.1 & \Sherpa 2.2.1 & Default & NNPDF3.0 & NLO \\
    Diboson (EW/loop)  & \Sherpa 2.1.1 & \Sherpa 2.1.1 & Default & CT10 & NLO \\
    \ttbar+$W/Z$  & \MGMCatNLO 2.2.3 & \PYTHIAV{8.212}  & A14 & NNPDF2.3 & NLO \\
    $\StopSym\StopSym^{*}$  & \MGMCatNLO 2.2.3 & \PYTHIAV{8.186} &  A14 & NNPDF2.3 & NLO+NLL~\cite{Borschensky:2014cia}\\

  \bottomrule
  \end{tabular}
}
\end{table}

The \ttbar and \singletop processes were simulated~\cite{ATL-PHYS-PUB-2016-004} at next-to-leading-order (NLO) accuracy
in perturbative QCD using the \POWHEGBOX~v2 event generator~\cite{Alioli:2010xd}
for \ttbar, \Wt, and $s$-channel \singletop production, and using the \POWHEGBOX~v1 generator for the electroweak $t$-channel \singletop production.
For these processes the spin correlations in top quark production and decay were preserved, and the top quark mass was set to 172.5~\GeV.
The matrix element was interfaced with the CT10 parton distribution function (PDF) set~\cite{CT10},
and the parton shower (PS), fragmentation, and underlying event were simulated with \PYTHIAV{6.428}~\cite{Sjostrand:2006za}
using the CTEQ6L1 PDF set~\cite{Pumplin:2002vw} and the P2012 underlying-event tuned parameters (UE tune)~\cite{Skands:2010ak},
with additional radiation simulated to the
leading-logarithm approximation through \pt-ordered parton showers~\cite{Corke2010}.

The \Zjet and \Wjet samples were generated at NLO~\cite{ATL-PHYS-PUB-2016-003} with the \Sherpa~2.2.1 event generator~\cite{sherpa}.
Matrix elements were calculated for up to two partons at NLO and four partons at LO using Comix~\cite{Gleisberg:2008fv} and OpenLoops~\cite{Cascioli:2011va},
and merged with the \Sherpa PS~\cite{Schumann:2007mg} using the ME+PS@NLO prescription~\cite{Hoeche:2012yf}.
The NNPDF3.0 PDF set~\cite{Ball:2014uwa} was used in conjunction with a dedicated PS tuning developed by the \Sherpa authors.
Diboson samples with two, three, or four leptons were similarly generated with \Sherpa~2.2.1.
The diboson matrix elements contain all diagrams with four electroweak vertices, and were calculated for up to one ($ZZ$) or zero ($WW$,$WZ$) partons at NLO and up to three partons at LO.
Electroweak- and loop-induced diboson events were simulated with \Sherpa~2.1.1, using the same prescriptions as above
but with the CT10 PDF set used in conjunction with the dedicated \Sherpa PS tuning.
The production of \ttbar with a $W$ or $Z$ boson (\ttV) was simulated at NLO using \MGMCatNLOFull (\MGMCatNLO)~2.2.3~\cite{Alwall:2014hca}
and interfaced to \PYTHIAV{8.212}~\cite{Sjostrand:2007gs} with the CKKW-L prescription~\cite{Lonnblad:2011xx}.
These samples are generated with the A14 UE tune~\cite{ATL-PHYS-PUB-2014-021} and NNPDF2.3 PDF set~\cite{Ball:2012cx}.

The RPV \Stop signal events were generated at leading order using the \MGMCatNLO~2.2.3 event generator with the NNPDF2.3 PDF set
and interfaced to \PYTHIAV{8.186}~\cite{Sjostrand:2007gs} using the A14 UE tune.
The matrix element was matched to the PS using the CKKW-L prescription, with the matching scale set to one quarter of the generated \Stop mass.
All other supersymmetric particles are assumed to be decoupled.
The signal cross sections are calculated to NLO accuracy in the strong coupling constant, adding the resummation of soft gluon emission at
next-to-leading-logarithm accuracy (NLO+NLL)~\cite{Kramer:2012bx,Beenakker:1997ut,Beenakker:2010nq,Beenakker:2011fu}.
The nominal cross section and the uncertainty for each mass value are taken from a combination of cross-section predictions
using different PDF sets and factorization and renormalization scales, as described in Ref.~\cite{Borschensky:2014cia}.
Stop samples were generated at masses between 600 and 1000~\GeV in steps of 100~\GeV
and between 1000 and 1600~\GeV in steps of 50~\GeV.
The cross section ranges from $ 175 \pm 23$~fb for a \StopSym mass of 600~\GeV to $0.141 \pm 0.038$~fb for a mass of 1600~\GeV.
The generated \Stops decay promptly through $\StopSym \rightarrow b\ell $ with a 1/3 branching ratio
($\mathcal{B}$) for each lepton flavor.
When optimizing the signal event selection, the generated events are reweighted to have $\BRe = \BRm = 0.5$ and $\BRt = 0$,
and various weightings are used to derive limits for different branching ratio assumptions.

All background samples are normalized using the available NLO or next-to-next-to-leading order (NNLO) cross sections, as indicated in \tabref{MC}.
The modeling of $c$-hadron and $b$-hadron decays in samples generated with \POWHEGBOX or \MGMCatNLO was performed with \textsc{EvtGen}~1.2.0~\cite{Lange:2001uf}.
Generated events were propagated through a full simulation of the ATLAS detector~\cite{SOFT-2010-01}
based on Geant4~\cite{GEANT4}, which describes the interactions of the particles with the detector.
A parameterized simulation of the ATLAS calorimeter called Atlfast-II~\cite{SOFT-2010-01}
was used for faster detector simulation of signal samples, and was found to agree well
with the full simulation.
Multiple overlapping $pp$ interactions (pileup) were included
by overlaying simulated minimum-bias events onto the simulated hard-scatter event.
Minimum-bias events were generated
using \PYTHIAV{8.186} with the A2 UE tune~\cite{ATL-PHYS-PUB-2012-003} and MSTW2008LO PDF set~\cite{Martin:2009iq}.
The simulated events are weighted such that the
distribution of the average number of $pp$ interactions per bunch crossing agrees with data.

\section{Event reconstruction}
\label{sec:reco}
Events and individual leptons and jets are required to satisfy several quality criteria to be considered by the analysis.
Events recorded during stable beam and detector conditions are required to satisfy data-quality criteria~\cite{Laycock:2253427}.
Each event is required to have a primary reconstructed vertex with two or more associated tracks with $\pt > 400~\MeV$,
where the primary vertex is chosen as the vertex with the highest $\Sigma \pt^{2}$ of associated tracks.
Two stages of quality and kinematic requirements are applied to leptons and jets.
The looser baseline requirements are first applied, and baseline leptons and jets are used to resolve
any misidentification or overlap between electrons, muons, and jets.
The subsequent tighter signal requirements are then applied to identify high-quality leptons and jets
in the kinematic phase space of interest.

Electron candidates are reconstructed from energy deposits in the
electromagnetic calorimeter matched to a charged-particle track in the ID.
Baseline electron candidates must have $\pt > 10~\GeV$, $|\eta| < 2.47$, and satisfy a loose
electron likelihood identification~\cite{ATL-PHYS-PUB-2016-015}.
Signal electrons must pass the baseline electron selection, have $\pt > 40~\GeV$,
and satisfy a tight electron likelihood identification.
In addition, they must be isolated from nearby activity, satisfying a loose \pt-dependent
track-based criterion~\cite{ATLAS-CONF-2016-024}.
Finally, their trajectory must be consistent with the primary vertex, such that their impact parameter
in the transverse plane ($d_0^{\mathrm{PV}}$) satisfies $|d_0^{\mathrm{PV}}|/\sigma_{d_0^{\mathrm{PV}}} < 5$, where $\sigma_{d_0^{\mathrm{PV}}}$
is the uncertainty in $d_0^{\mathrm{PV}}$.
Each signal electron must have a longitudinal impact parameter with respect to the primary vertex ($z_0^{\mathrm{PV}}$)
that satisfies $|z_0^{\mathrm{PV}}~\text{sin}\theta| < 0.5$~mm.

Muon candidates are reconstructed by combining tracks in the ID with tracks in the MS.
Baseline muon candidates must have \pt $>10~\GeV$, $|\eta| < 2.7$, and satisfy the
medium muon identification criteria~\cite{PERF-2015-10}.
Signal muons must pass the baseline muon selection, have $\pt > 40~\GeV$, $|\eta| < 2.5$,
$|z_0^{\mathrm{PV}}~\text{sin}\theta| < 0.5$~mm, and $|d_0^{\mathrm{PV}}|/\sigma_{d_0^{\mathrm{PV}}} < 3$.
As with electrons, muons must satisfy the \pt-dependent loose track-based isolation criteria.
Events containing a poorly measured signal muon, as determined by having
incompatible momentum measurements in the ID and the MS, are rejected.
Absolute requirements of $|z_0^{\mathrm{PV}}| < 1$~mm and $|d_0^{\mathrm{PV}}| < 0.2$~mm on the impact parameters of signal muons are applied to reject cosmic muons.

Jets are reconstructed using the anti-$k_{t}$ algorithm~\cite{Cacciari:2008gp,Cacciari:2011ma} with a radius
parameter $R = 0.4$ from clusters of energy deposits in the calorimeters~\cite{PERF-2014-07}.
Jets are corrected for pileup contamination on an event-by-event basis using the jet area subtraction method~\cite{PERF-2014-03,Cacciari:2007fd}.
Jets are further calibrated to account for the predicted detector response in MC simulation,
and a residual calibration of jets in data is derived through \textit{in situ} measurements~\cite{Aaboud:2017jcu}.
Baseline jet candidates are required to have $\pt >20~\GeV$ and $|\eta| < 2.8$.
Jets with $\pt < 60~\GeV$ and $|\eta| < 2.4$ are required to satisfy pileup-rejection criteria based on charged-particle tracks and implemented through the jet vertex tagger algorithm~\cite{PERF-2014-03}.
Signal jets must pass the baseline jet selection and have $\pt > 60~\GeV$.
Events are rejected if they contain a jet that fails the loose quality criteria~\cite{ATLAS-CONF-2015-029}, reducing contamination from calorimeter noise bursts and non-collision backgrounds.
Jets within $|\eta| < 2.5$ that are initiated by $b$-quarks are identified using the multivariate \texttt{MV2c10} $b$-tagging algorithm~\cite{PERF-2012-04,ATL-PHYS-PUB-2016-012},
which exploits the impact parameters of charged-particle tracks,
the parameters of reconstructed secondary vertices, and the topology of $b$- and $c$-hadron decays inside a jet.
The working point is chosen to provide a $b$-tagging efficiency of 77\% per $b$-jet in simulated $\ttbar$ events
with a rejection factor of approximately 130 for jets initiated by gluons or light-flavor quarks and 6 for jets initiated by
$c$-quarks~\cite{ATL-PHYS-PUB-2016-012}.
Correction factors are applied to events to compensate for differences between data and MC simulation in the $b$-tagging efficiency for $b$-jets, $c$-jets, and light-flavor jets.

To avoid reconstructing a single detector signature as multiple leptons or jets, an overlap removal
procedure is performed on baseline leptons and jets.
The requirements are applied sequentially, and failing particles are removed from consideration in the subsequent steps.
If an electron and muon share a track in the ID, the electron is removed.
Any jet that is not $b$-tagged and is within a distance\footnote{
  The distance between two four-momenta is defined as $\Delta R = \sqrt{(\Delta y)^{2}+(\Delta\phi)^{2}}$,
  where $\Delta y$ is their distance in rapidity and $\Delta\phi$ is their azimuthal distance.
  The distance with respect to a jet is calculated from its central axis.}
$\Delta R(\ell,\mathrm{jet}) \le 0.2$ of a lepton is removed.
If the jet is $b$-tagged, the lepton is removed instead in order to suppress leptons from semileptonic decays of $c$- and $b$-hadrons.
Finally, any lepton that is $\Delta R(\ell,\mathrm{jet}) \le 0.4$ from a jet is removed.

The trigger, reconstruction, identification, and isolation efficiencies of electrons~\cite{ATLAS-CONF-2016-024} and muons~\cite{PERF-2015-10}
in MC simulation are corrected using events in data with leptonic $Z$ and $\Jpsi$ decays.
Similarly, corrections to the $b$-tagging efficiency and mis-tag rate in MC simulation are derived from various control regions in data~\cite{ATL-PHYS-PUB-2016-012}.

\section{Event selection}
\label{sec:sel}
To identify the pair production of \Stops, events are required to have at least two leptons and two jets.
If more than two leptons or two jets are found, the two highest-\pt leptons and jets are selected.
At least one of the two leading jets must be $b$-tagged.
The selected leptons are required to have opposite charge, and one of them must be consistent with the associated single-lepton trigger.
This trigger requirement is highly efficient for signal events, with an efficiency of 93\% for the $\mu\mu$ channel, 95\% for the $e\mu$ channel,  and 98\% for the $ee$ channel.

The lepton--jet pair from each \StopSym decay generally reconstructs the invariant mass $\Mbl$ of the original \StopSym.
In an event with two leptons and two jets, two pairings are possible; one that reconstructs the correct
\StopSym masses, and one which inverts the pairing and incorrectly reconstructs the masses.
As the two masses should be roughly equal, the pairing that minimizes the mass asymmetry between $\Mbl^0$ and $\Mbl^1$ is chosen, defined as

\begin{equation*}
  \MblAsym = \frac{\Mbl^0-\Mbl^1}{\Mbl^0+\Mbl^1}.
\end{equation*}

Here $\Mbl^0$ is chosen to be the larger of the two masses.
Events are further selected to have small mass asymmetry $\MblAsym < 0.2$.
This reduces the contamination from background processes, whose random pairings lead to a more uniform $\MblAsym$ distribution.

Two nested signal regions (SRs) are constructed to optimize the identification of signal over background events.
The signal regions are optimized using MC signal and background predictions, assuming \StopSym decays of \BRe = \BRm = 50\%.
A primary kinematic selection of the signal regions is on $\Mbl^0$, with SR800 requiring $\Mbl^0 > 800~\GeV$ and SR1100 requiring $\Mbl^0 > 1100~\GeV$.
By defining two signal regions the sensitivity to high-mass signals above 1100~\GeV is improved, while maintaining sensitivity to lower-mass signals.
Several other kinematic selections, common to both SRs, are defined to reduce the contribution from the largest backgrounds.
As the \StopSym decay products are generally very energetic, a selection on their \pt sum,

\begin{equation*}
 \Ht = \sum_{i=1}^{2}\pt^{\ell_i} + \sum_{j=1}^{2}\pt^{\mathrm{jet}_j}
 \label{eq:Ht}
 \end{equation*}

is applied, such that $\Ht > 1000~\GeV$.
To reduce contamination from \Zjet events, a requirement is placed on the invariant mass of two same-flavor leptons, with $\mll > 300~\GeV$.
A large fraction of the background from processes involving a top quark is suppressed through the requirement on $\Mbl^0$ and $\MblAsym$,
with correctly reconstructed top quark masses falling well below the signal region requirements.
However, top quark decays in which the lepton and $b$-jet decay products are mispaired can enter the SRs if the incorrectly reconstructed masses happen to be large.
In such cases it is the rejected pairing that properly reconstructs the top quark decay, with one of the two \bl pair masses
below the kinematic limit for a top quark decay.
To suppress such backgrounds, events are rejected if the subleading \bl mass of the rejected pairing, \mblrej,
is compatible with that of a reconstructed top quark, with $\mblrej < 150~\GeV$.

The distribution of predicted signal and background events is shown for the SR800 region in \figref{N1} for $\Mbl^0$, \Ht, \MblAsym, \mll, and \mblrej,
demonstrating the potential for background rejection.
For the model with a \StopSym mass of 1000~\GeV (1500~\GeV), the SR800 selections are 21\% (24\%) efficient for events with two $\StopSym~\to~be$ decays,
16\% (16\%) for events with two $\StopSym~\to~b\mu$ decays, and 0.1\% (0.3\%) for events with two $\StopSym~\to~b\tau$ decays.

\begin{figure}[!h]
  \centering
  \subfigure[$\Mbl^0$]{
    \includegraphics[width=0.47\textwidth]{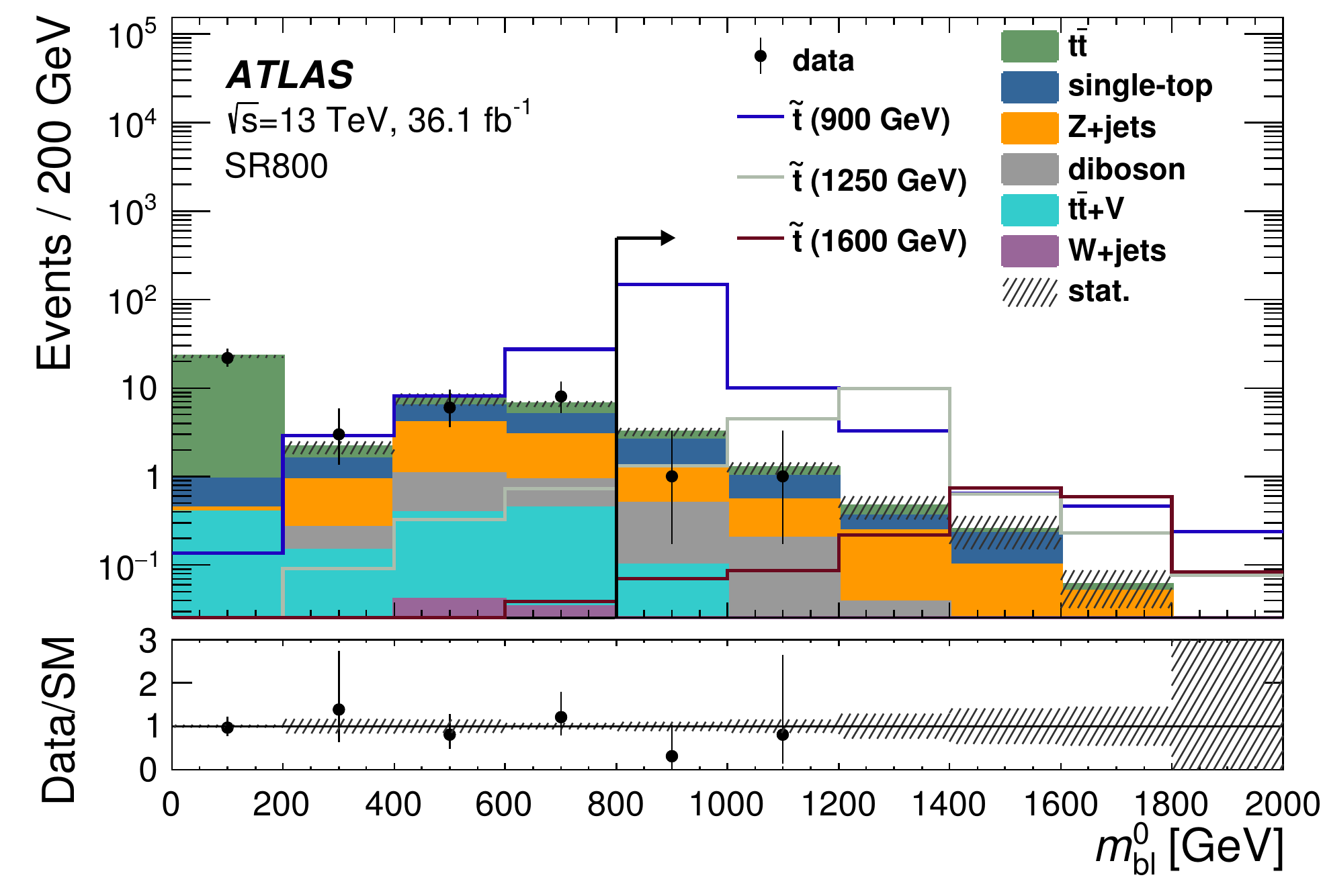}
    \label{fig:N1:mbl0}
  }
    \subfigure[\MblAsym]{
    \includegraphics[width=0.47\textwidth]{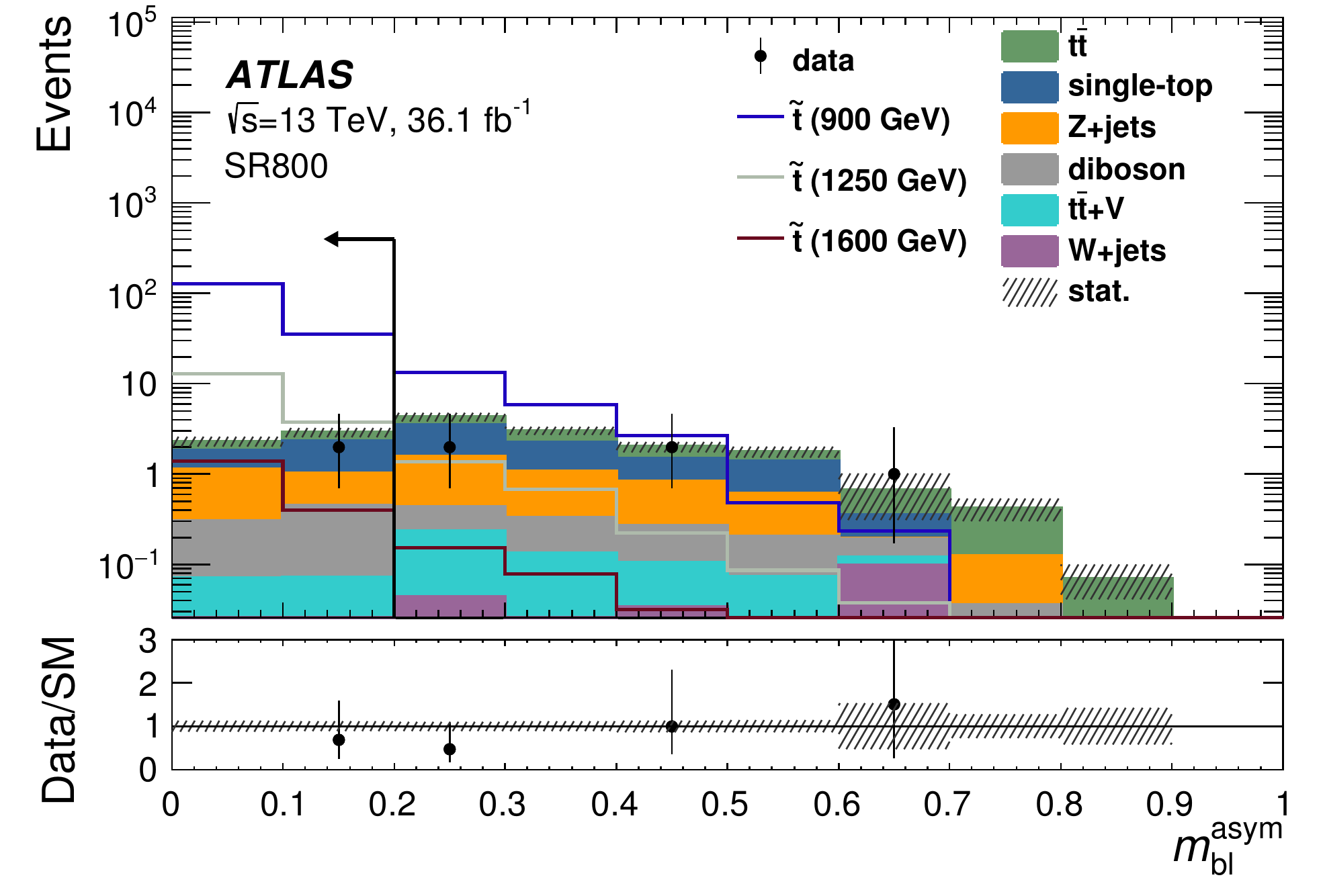}
    \label{fig:N1:mblAsym}
  }
    \subfigure[\Ht]{
    \includegraphics[width=0.47\textwidth]{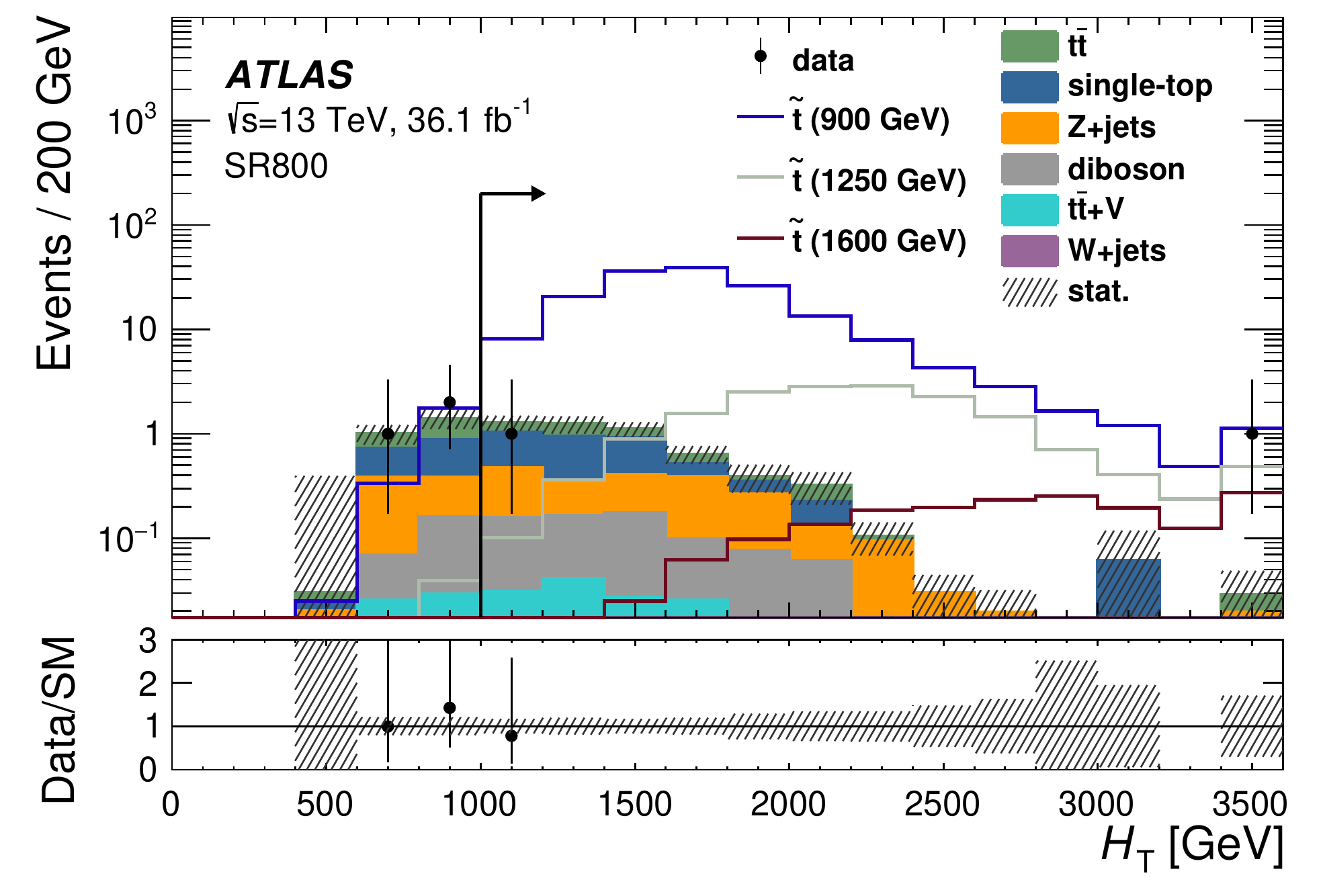}
    \label{fig:N1:HT}
  }
    \subfigure[\Mll]{
    \includegraphics[width=0.47\textwidth]{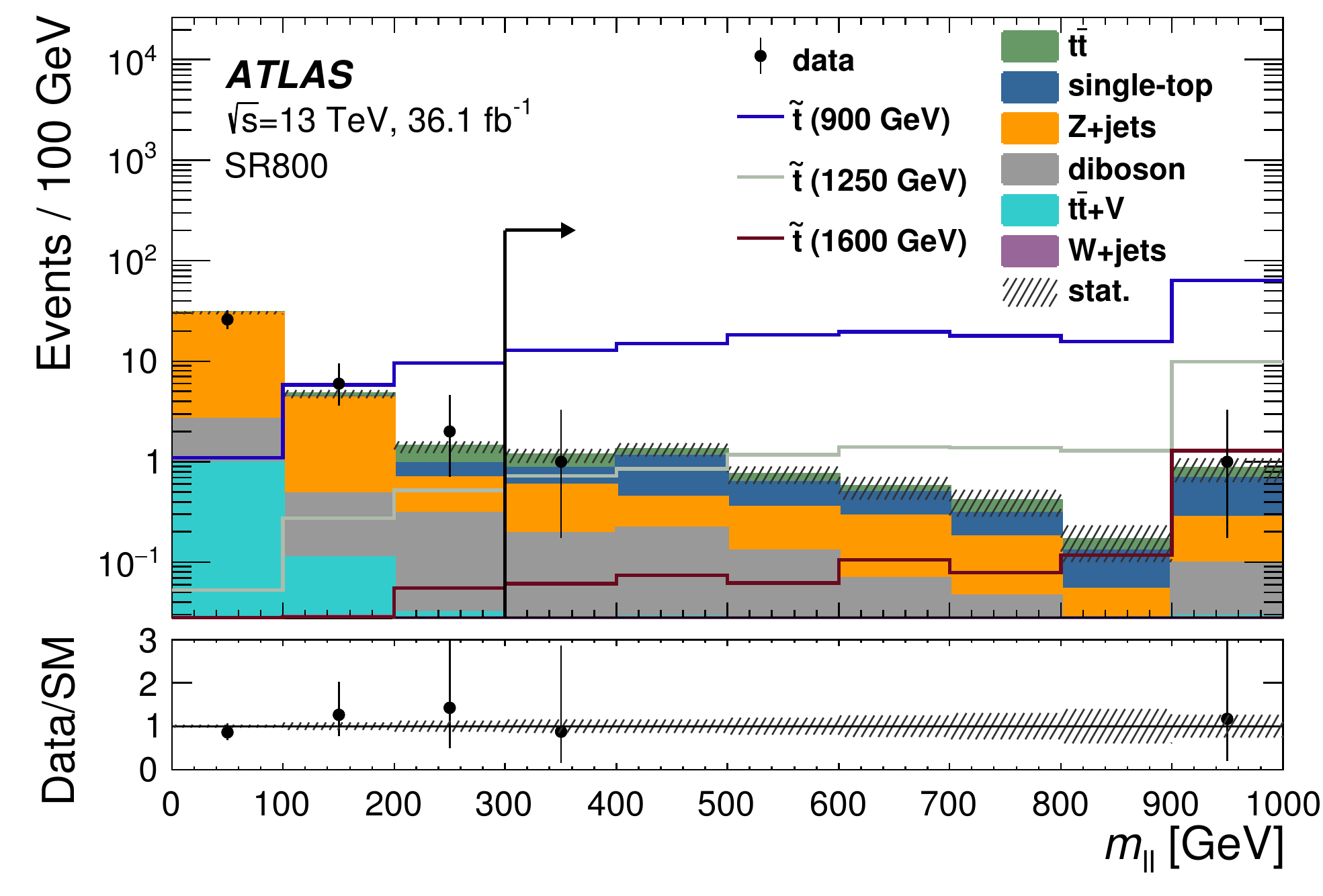}
    \label{fig:N1:mll}
  }
    \subfigure[$\mblrej$]{
    \includegraphics[width=0.47\textwidth]{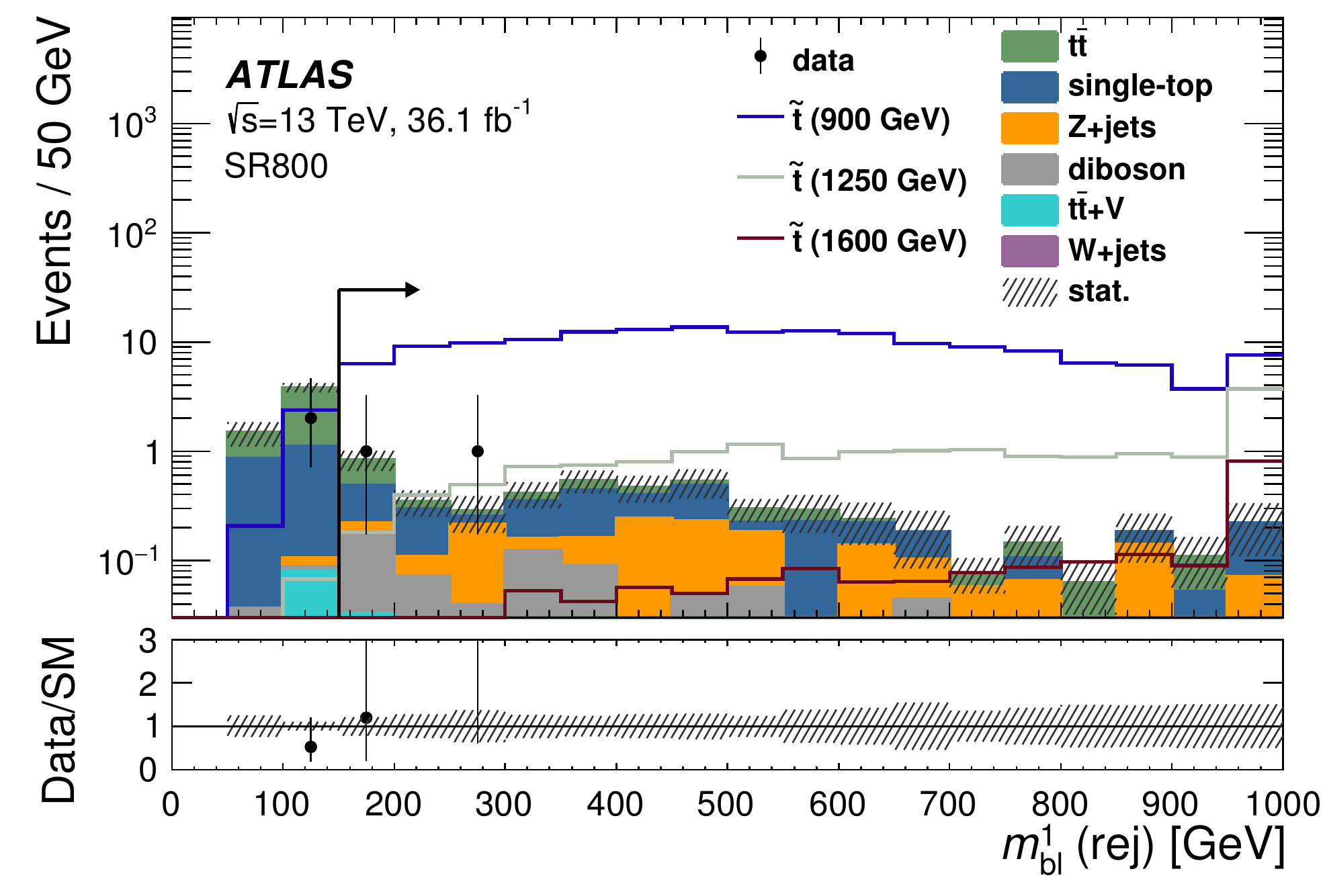}
    \label{fig:N1:mblrej1}
  }

  \caption{ Distributions of \subref{fig:N1:mbl0} $\mbl^{0}$,
  \subref{fig:N1:mblAsym} \MblAsym, \subref{fig:N1:HT} \Ht,
  \subref{fig:N1:mll} \mll, and \subref{fig:N1:mblrej1} $\mblrej$ in the SR800 signal region
  for the data and post-fit MC prediction.
  The SR800 event selections are applied for each distribution except the selection on the variable shown, which is indicated by an arrow.
  Normalization factors are derived from the background-only estimation discussed in \secref{bkg} and are applied to the dominant
  \ttbar, \singletop, and \Zjet processes.
  Benchmark signal models generated with \StopSym masses of 900, 1250, and 1600~\GeV are included for comparison.
  The bottom panel shows the ratio between the data and the post-fit MC prediction.
  The hatched uncertainty band includes the statistical uncertainties in the background prediction.
  The last bin includes the overflow events.
  }
  \label{fig:N1}
\end{figure}

\section{Background estimation}
\label{sec:bkg}

For each of the relevant backgrounds in the signal regions, one of two
methods is used to estimate the contribution.
The minor backgrounds from \diboson, \ttV, and \Wjet processes are estimated directly from MC simulation
and the normalization is corrected to the highest-order theoretical cross section available.
For the dominant \ttbar, \singleTop, and \Zjet backgrounds, the expected yield in the SRs
is estimated by scaling each MC prediction by a normalization factor
(NF) derived from three dedicated control regions (CRs), one for each
background process.  Each control region is defined to be
kinematically close to the SRs while inverting or relaxing specific
selections to enhance the contribution from the targeted background
process while reducing the contamination from other backgrounds and
the benchmark signals.

To derive a background-only estimate, the normalizations of
the \ttbar, \singleTop, and \Zjet backgrounds are determined through a
likelihood fit~\cite{Baak:2014wma} performed simultaneously to the
observed number of events in each CR.  The expected yield in each
region is given by the inclusive sum over all background processes in
the \EE, \EM, and \MM channels.  The NF for each of
the \ttbar, \singleTop, and \Zjet backgrounds are free parameters of the fit. The systematic
uncertainties are treated as nuisance parameters in the fit and are not significantly constrained.

Several validation regions (VRs) are defined to test the extrapolation from the CRs to SRs over the relevant kinematic variables.
The VRs are disjoint from both the CRs and SRs, and are constructed to fall between one or more CRs and the SRs
in one of the extrapolated variables.
The VRs are not included in the fit, but provide a statistically independent cross-check of the background prediction
in regions with a negligible signal contamination.
Three VRs are constructed to test the extrapolation in the $\mbl^0$, \Mblrej, and \Ht observables.
A fourth VR is constructed to validate the extrapolation of the \Zjet CR in \mll.
Details of the selection criteria in each CR and VR are presented below, and a summary of the selections is provided in \tabref{region_cuts}.

\begin{table}[t]
  \centering{
  \caption{
    Summary of the selections of the signal, control, and validation regions.
    All regions require at least two oppositely charged leptons and at least two jets.
    Each region requires at least one of the two leading jets to be $b$-tagged with the exception of CRst,
    which requires both leading jets to be $b$-tagged, and \VRZ, which requires zero $b$-tagged jets in the event.
    A mass asymmetry selection of $\MblAsym < 0.2$ is applied to all regions.
    The contransverse mass selection \mct (\equref{mct}) is only applied to events in
    CRtt with exactly two $b$-tagged jets, as indicated by the ${}^*$, ensuring the region is orthogonal to CRst.
  }
  \label{tab:region_cuts}
    \begin{tabular}{c|ccccccc}
      \toprule
      Region & $N_{b}$ & $\Mbl^0$ [GeV] & \Ht [GeV] & $\Mblrej$[GeV] & \mll [GeV] & \mct [GeV] \\
      \hline
      \hline
      SR800   & $\ge 1$ & $> 800$  & $> 1000$   & $> 150$ & $>300$& --  \\
      SR1100   & $\ge 1$ & $> 1100$  & $> 1000$ & $> 150$ & $>300$ & --  \\
      \hline
      CRst   & $=2$ & [$200$,$500$]  & $<800$ & $<150$  & $>120$ & $> 200$  \\
      CRtt   & $\ge 1$ & [$200$,$500$]  & [$600$,$800$] & $<150$  & $>300$ & $<200^*$ \\
      CRZ   & $\ge 1$ &  $>700$  & $>1000$ & -- &  [$76.2$,$106.2$] & -- \\
      \hline
      \VRacc & $\ge 1$ & $>500$ & [$600$,$800$] & $<150$  & $>300$ & -- \\
      \VRrej & $\ge 1$ & [$200$,$500$]  & [$600$,$800$] & $>150$  &  $>300$ & -- \\
      \VRht & $\ge 1$ & [$200$,$500$]  & $>800$ & $<150$  & $>300$ & -- \\
      \VRZ  & $=0$ & [$500$,$800$] & $>1000$ & $>150$ & $>300$ & -- \\
      \bottomrule
    \end{tabular}
  }
\end{table}

\subsection{\Singletop control region}
\label{sec:CRst}
The \singletop background enters the SR through the \Wt process, when the $b$-jet and lepton produced in the semileptonic top quark decay
are incorrectly paired with the lepton from the $W$ decay and an additional jet, respectively.
The CRst control region is designed to target the \Wt production in a less-energetic kinematic region or
where the rejected \bl pairing correctly combines the decay products of the top quark.
To separate CRst from the SRs, the \Ht and $\Mbl^0$ requirements are reversed such that $\Ht < 800~\GeV$ and
$200 < \Mbl^0 < 500~\GeV$.
To target events in which the top quark is reconstructed in the rejected \bl pairing, the selection on \Mblrej is reversed, requiring $\Mblrej < 150~\GeV$.
As there is no dilepton resonance in this background process the \mll selection is lowered to
increase the CRst yield and improve the statistical precision of the constraint.

After these selections the control region is dominated by \ttbar production, which has a significantly higher cross section than the \Wt process.
The contransverse mass (\mct)~\cite{Tovey:2008ui} is introduced to discriminate between \Wt and \ttbar events and increase the \Wt purity in the CRst.
The \mct observable attempts to reconstruct the invariant mass of pair-produced particles which decay into visible and invisible decay products.
For two identical decays of top quarks into two visible $b$-quarks $b_{1}$ and $b_{2}$,
and two W bosons, each of whose decay products may include an invisible particle, \mct is defined as
\begin{equation}
\mct^{2}(b_1, b_2) = [\et(b_1)+\et(b_2)]^{2} - [\mathbf{p}_{\mathrm T}(b_1) - \mathbf{p}_{\mathrm T}(b_2)]^{2},
 \label{eq:mct}
\end{equation}

\noindent where $\et = \sqrt{\pt^{2}+m^{2}}$ is calculated from the kinematics of the reconstructed $b$-jet.
For an event with two top quarks, the \mct observable therefore has a kinematic endpoint at
\begin{equation*}
\mct^{\mathrm{max}} = \frac{m_{t}^2 - m_{W}^2}{m_{t}},
\end{equation*}

where $m_{t}$ and $m_{W}$ are the masses of the top quark and $W$ boson, respectively.
Requiring this variable to exceed a minimum value is effective in suppressing the \ttbar contribution,
for which \mct has a kinematic endpoint of about 135~\GeV,
and a strict requirement of $\mct > 200~\GeV$ is applied in CRst.
The \mct variable is only effective in rejecting \ttbar events in which the $b$-quark decay products of both top quarks
are properly identified, and both leading jets (and only the leading jets) are required to be $b$-tagged in CRst, such that $N_b=2$.
The \mct distribution of the backgrounds in CRst is shown in \figref{CR_distributions:mct} when no \mct requirement is applied, and a significant
\singletop contribution above $55\%$ is seen for $\mct > 200~\GeV$.

\subsection{\ttbar control region}
The CRtt control region is constructed to target \ttbar events with kinematics similar to the SRs.
As with CRst, the \Ht and $\Mbl^0$ requirements are inverted such that $600 < \Ht < 800~\GeV$ and $200 < \Mbl^0 < 500~\GeV$.
The selection on \Mblrej is also inverted, requiring $\Mblrej < 150~\GeV$, such that one of the two top quarks
is reconstructed in the rejected \bl pairings.
The distribution of \Mblrej in CRtt is shown in \figref{CR_distributions:mblrej},
showing the mispairing of \ttbar events is well-modeled in MC simulation.
Due to the larger cross section of the \ttbar process, contamination from \Wt events is minimal.
However, to maintain orthogonality with CRst, a requirement of $\mct < 200~\GeV$ is applied to events in which
both leading jets (and only the leading jets) are $b$-tagged, with $N_b=2$.

\subsection{\Zjet control region}
The CRZ control region targets \Zjet events by applying a selection on the invariant mass of the dilepton pair \mll,
requiring it to be within 15~\GeV of the $Z$ mass.
Both leptons are required to be of the same flavor.
The \mll selection is effective in removing signal contamination, and the SR \Ht selection is left unchanged, while the
$\mbl^0$ selection is slightly relaxed to $\mbl^0 > 700~\GeV$ to enchance the event yield.

\begin{figure}[h]
  \centering
  \subfigure[\mct]{
    \includegraphics[width=0.47\textwidth]{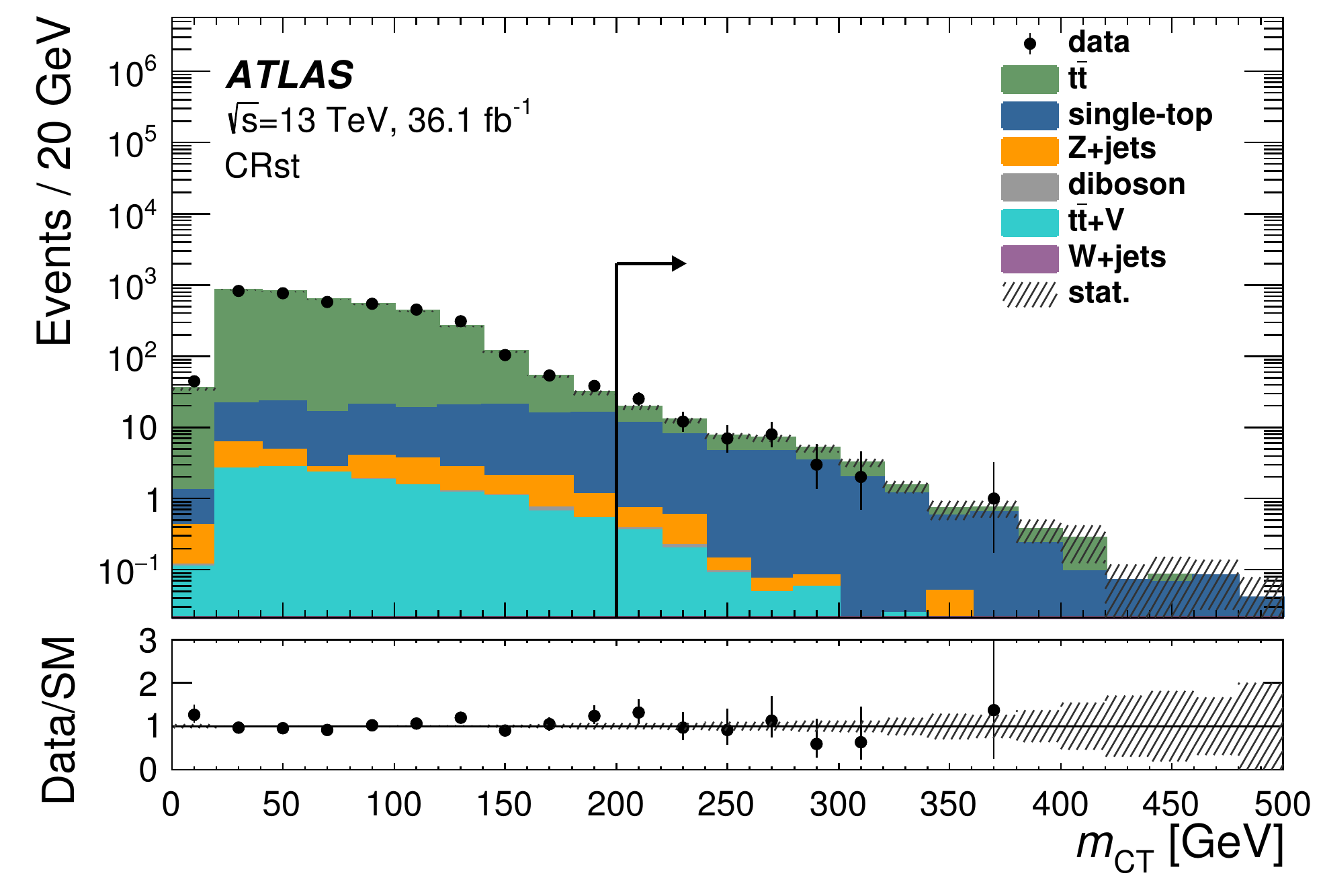}
    \label{fig:CR_distributions:mct}
  }
  \subfigure[\Mblrej]{
    \includegraphics[width=0.47\textwidth]{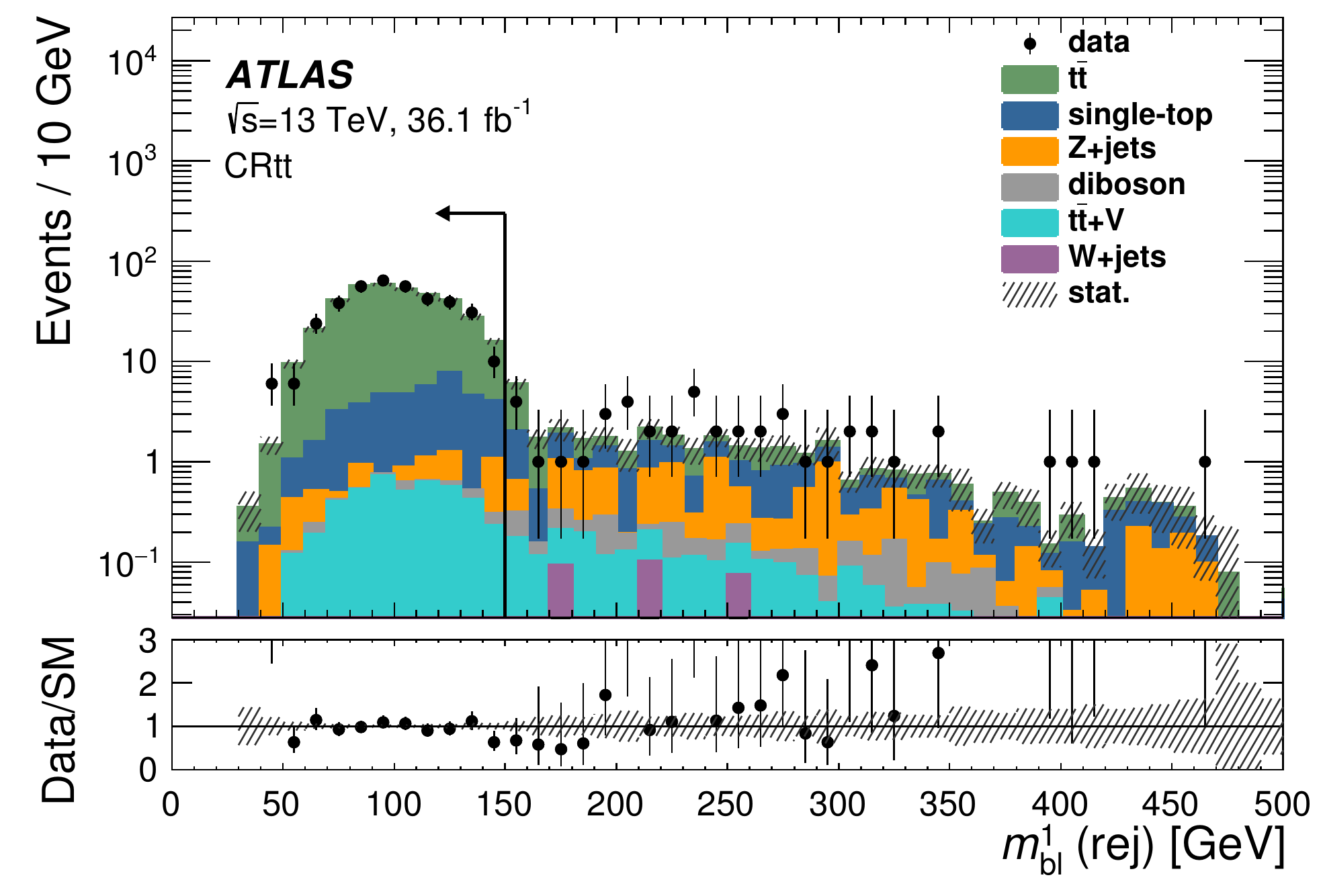}
    \label{fig:CR_distributions:mblrej}
  }
  \caption{
    Distributions of \subref{fig:CR_distributions:mct} \mct in CRst and \subref{fig:CR_distributions:mblrej} \Mblrej in CRtt for the data and post-fit MC prediction.
    The relevant CR event selections are applied for each distribution except the selection on the variable shown, which is indicated by an arrow.
    Normalization factors are derived from the background-only fit configuration and are applied to the dominant
    \ttbar, \singletop, and \Zjet processes.
    The bottom panel shows the ratio between the data and the post-fit MC prediction.
    The hatched uncertainty band includes the statistical uncertainties in the background prediction.
    The last bin includes the overflow events.
  }
  \label{fig:CR_distributions}
\end{figure}

\subsection{Validation regions}
Four disjoint validation regions are used to test the extrapolation of the background fit from the CRs to the SRs.
A full list of the region selections is given in \tabref{region_cuts}.
The \VRacc, \VRrej, and \VRht test the extrapolation from CRst and CRtt to the SRs in the $\Mbl^0$, \Mblrej, and \Ht observables
by requiring $\Mblacc > 500~\GeV$, $\Mblrej > 150~\GeV$, and $\Ht > 800~\GeV$, respectively.
In this way \VRacc, \VRrej, and \VRht all lie between the SRs and both CRtt and CRst, with signal contamination below 1\% for all signal mass values.
No requirement is placed on $\mct$ in any VR, allowing both the \ttbar and \Wt contributions to be validated.

A fourth validation region, \VRZ, is used to test the extrapolation from CRZ to the SRs in the \mll observable, requiring $\mll > 300~\GeV$.
As the \mll variable provides the only separation between CRZ and the SRs, the requirement on \Mblacc is relaxed to $500 < \Mbl^0 < 800~\GeV$,
and any event with a $b$-tagged jet is rejected, such that $N_b=0$.
The \Zjets MC prediction is found to model the data well in both $\Mbl$ and $N_b$, with a signal contamination in \VRZ
below 5\% for mass values above 1000~\GeV.

The observed data yield and the post-fit background prediction for each CR and VR are shown in \figref{regionYields}.
Good agreement is seen in all validation regions, with differences between the data and SM prediction below $1\sigma$.
The modeling of the extrapolated variable for each VR is shown in \figref{VR_distributions}, demonstrating good agreement in the shape
of the variables of interest.

\begin{figure}[h]
  \centering
  \includegraphics[width=0.85\textwidth]{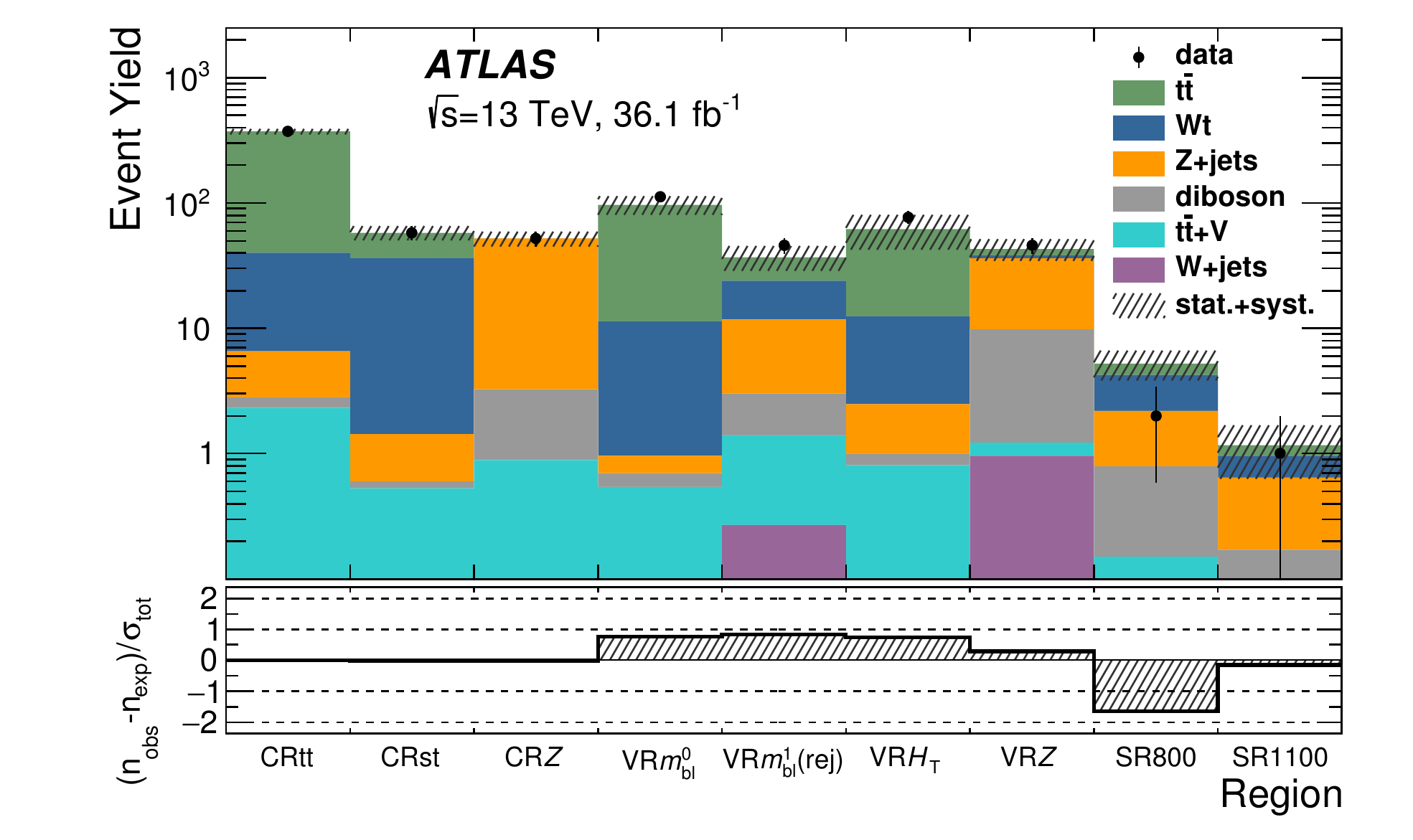}
  \caption{
    Comparison of the observed data and expected numbers of events in the CRs, VRs, and SRs.
    The background prediction is derived with the background-only fit configuration, and the
    hatched band includes the total uncertainty in the background prediction.
    The bottom panel shows the significance of the difference between data and the background prediction.
  }
  \label{fig:regionYields}
\end{figure}

\begin{figure}[h]
  \centering
  \subfigure[\Mblacc]{
    \includegraphics[width=0.47\textwidth]{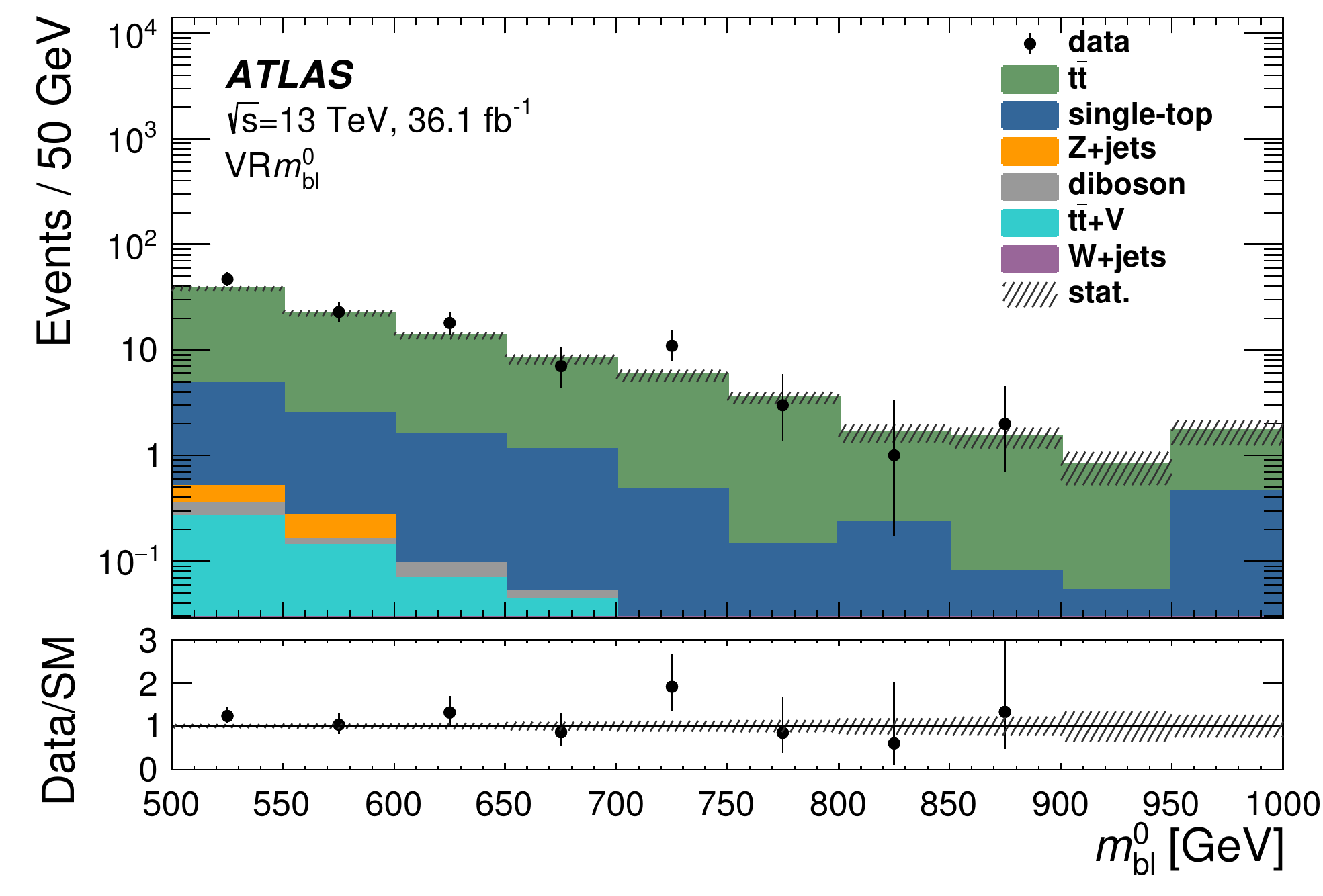}
    \label{fig:VR_distributions:mbl0}
  }
  \subfigure[\Mblrej]{
    \includegraphics[width=0.47\textwidth]{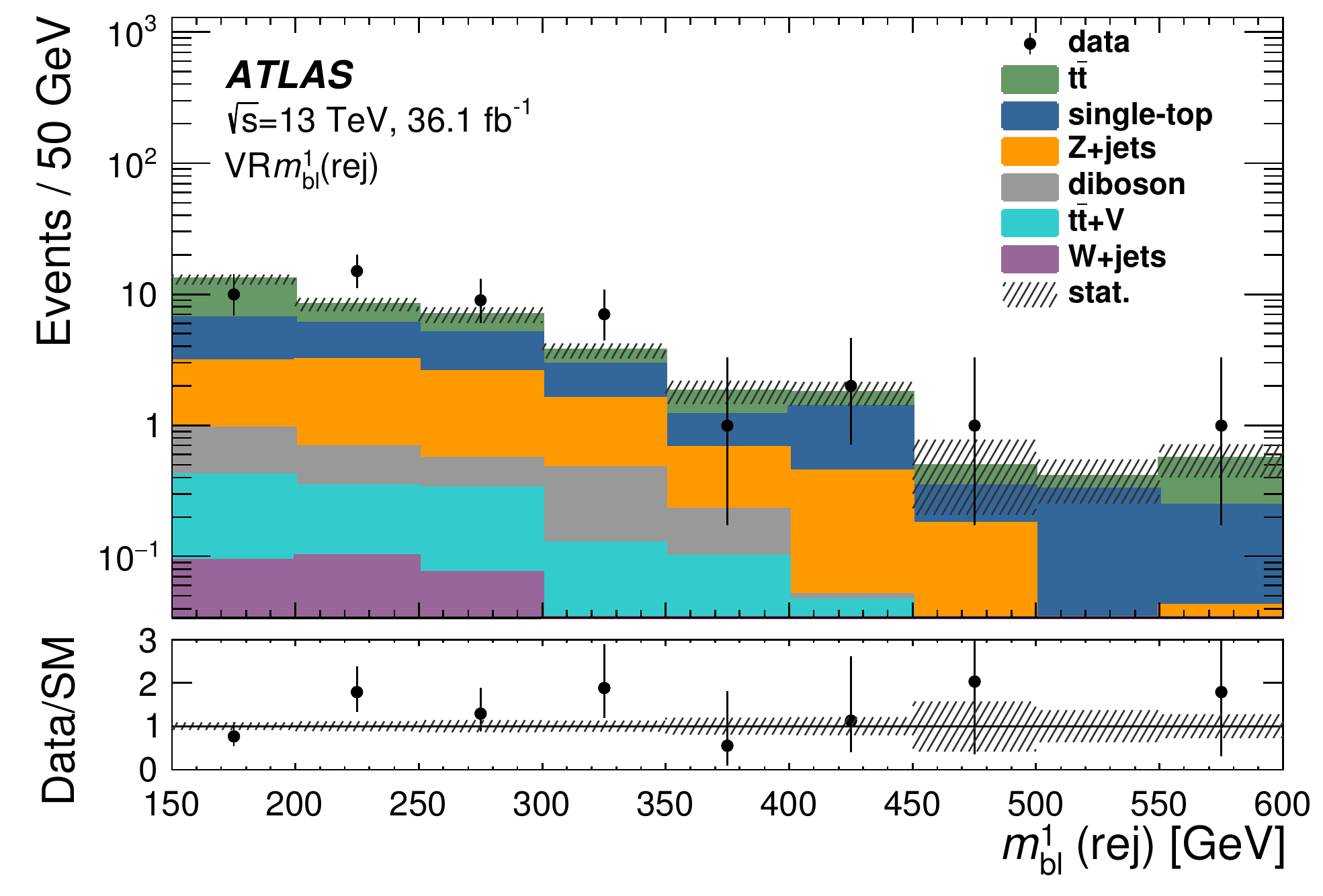}
    \label{fig:VR_distributions:mblrej}
  }
  \subfigure[\Ht]{
    \includegraphics[width=0.47\textwidth]{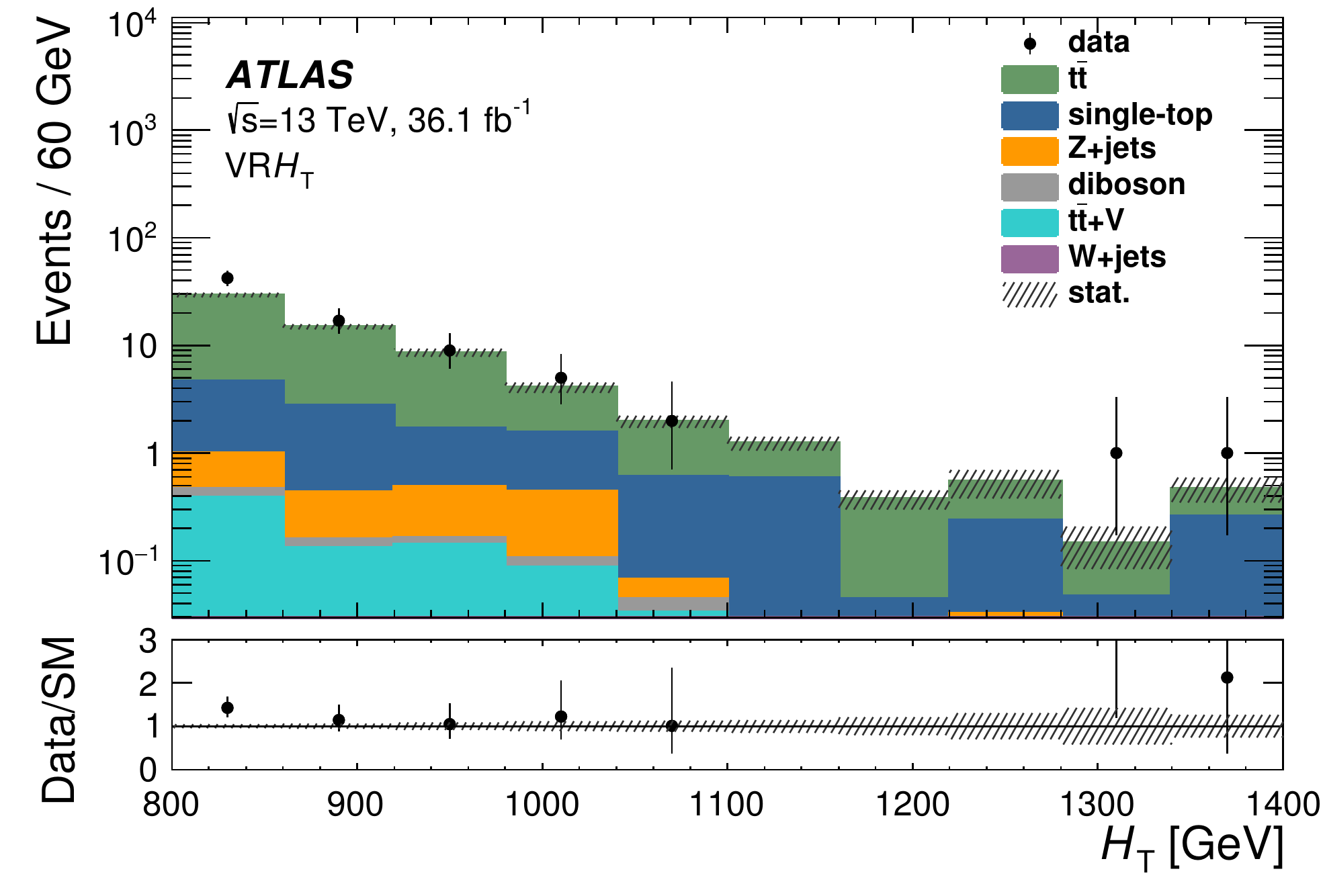}
    \label{fig:VR_distributions:ht}
  }
  \subfigure[\mll]{
    \includegraphics[width=0.47\textwidth]{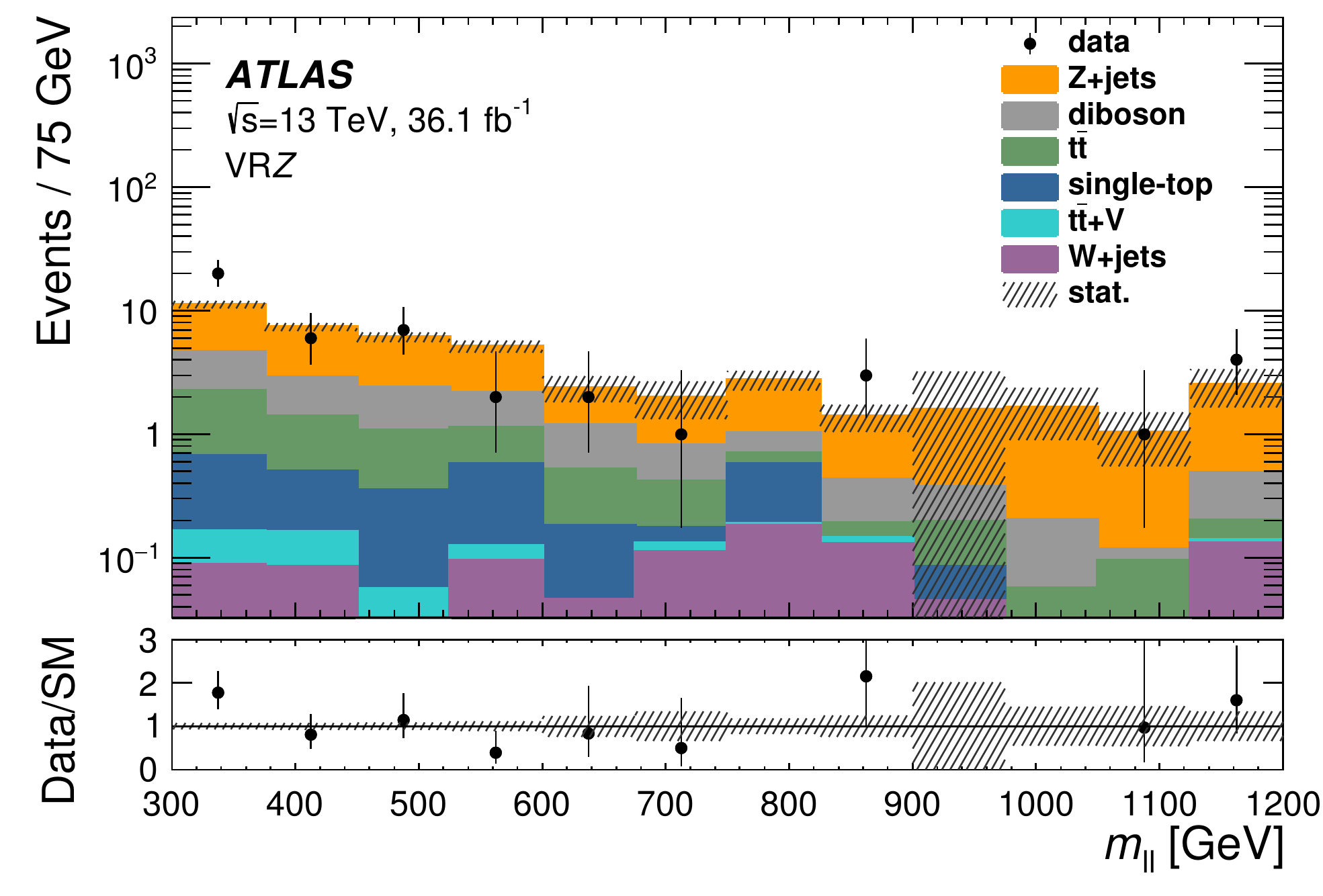}
    \label{fig:VR_distributions:mll}
  }
  \caption{
    Distributions of \subref{fig:VR_distributions:mbl0} \Mblacc in \VRacc, \subref{fig:VR_distributions:mblrej} \Mblrej in \VRrej for the data and post-fit MC prediction,
    \subref{fig:VR_distributions:ht} \Ht in \VRht, and \subref{fig:VR_distributions:mll} \mll in \VRZ.
    Normalization factors are derived from the background-only fit configuration and are applied to the dominant
    \ttbar, \singletop, and \Zjet processes.
    The bottom panel shows the ratio between the data and the post-fit MC prediction.
    The hatched uncertainty band includes the statistical uncertainties in the background prediction.
    The last bin includes the overflow events.
  }
  \label{fig:VR_distributions}
\end{figure}

\section{Systematic uncertainties}
\label{sec:systematics}
Systematic uncertainties in the signal and background predictions arise
from theoretical uncertainties in the expected yield and MC modeling, and from experimental sources.
The dominant uncertainties are summarized in \tabref{syst_summary}.

\begin{table}[t]
\begin{center}
  \caption{Summary of the dominant experimental and theoretical uncertainties in SR800 and SR1100 before the likelihood fits,
   quoted relative to the total pre-fit MC background predictions.
  The individual uncertainties can be correlated, and do not necessarily add in quadrature to the total post-fit background uncertainty.}
  \label{tab:syst_summary}
\setlength{\tabcolsep}{0.0pc}
{\small
  \begin{tabular*}{\textwidth}{@{\extracolsep{\fill}}l rr}
  \midrule
  \textbf{Source \textbackslash~Region}    &     SR800    & SR1100    \\[-0.05cm]
  \midrule
  \midrule
  Experimental uncertainty  &  & \\
  \midrule
  $b$-tagging & $3\%$ & $5\%$  \\
  Jet energy resolution  & $2\%$ & $10\%$  \\
  Jet energy scale  & $1\%$ & $3\%$  \\
  Electrons &  $1\%$  &  $4\%$ \\
  Muons  & $1\%$ & $3\%$ \\
  \midrule
  Theoretical modeling uncertainty  &  &  \\
  \midrule
  MC statistical uncertainty &  $8\%$ &  $17\%$  \\
  \ttbar & $8\%$ & $45\%$  \\
  Single-top & $21\%$  & $22\%$ \\
  \Zjet & $2\%$ & $4\%$  \\
  Diboson & $4\%$ & $3\%$ \\
  \ttbar+$W/Z$ & 1\% & 1\% \\
  \Wjet & 1\% & 1\% \\

  \noalign{\smallskip}\hline\noalign{\smallskip}
  \end{tabular*}
  }
  \end{center}
  \vspace{-0.2cm}
\end{table}

Experimental uncertainties reflect the precision of the energy and momentum calibration of jets and leptons, as well as the
assumptions about the identification and reconstruction efficiencies in MC simulation.
The dominant experimental uncertainties are related to jets, including those in the jet energy scale and resolution~\cite{PERF-2011-04,ATL-PHYS-PUB-2015-015}
and the calibration of the $b$-tagging efficiency for $b$-jets, $c$-jets, and light-flavor jets~\cite{ATL-PHYS-PUB-2016-012}.
The largest experimental uncertainties in the fitted background prediction in SR800 (SR1100) are from the
$b$-tagging efficiency of light-flavor jets and the jet energy resolution.
The experimental uncertainties associated with leptons each have a small impact on the final measurement, and include uncertainties in
the energy scale and resolution of electrons~\cite{ATLAS-CONF-2016-024} and muons~\cite{PERF-2015-10},
and the calibration of the lepton trigger, identification, reconstruction, and isolation efficiencies.
The 3.2\% uncertainty in the measured integrated luminosity also has a marginal effect on the final result.

Theoretical and MC modeling uncertainties of the \ttbar and \Wt backgrounds account for the choice of event generator, underlying-event tune, and their parameters.
The uncertainties are derived separately for each background process and are treated as uncorrelated nuisance parameters.
As the \ttbar (\Wt) background normalization is constrained in the likelihood fits, the uncertainties are derived on the transfer of the NF from the CRtt (CRst)
to both SR800 and SR1100 by comparing CR-to-SR yield ratios in alternative models.
The uncertainty in the background estimate due to the choice of MC event generator is estimated for \ttbar and \Wt
by comparing the CR-to-SR yield ratios derived using \MGMCatNLO~2.2.3 with the one derived using \POWHEGBOX~v2,
both showered with \Herwigpp~v2.7.1~\cite{Marchesini:1991ch} using the UEEE5 UE tune~\cite{Gieseke:2012ft}.
The generator uncertainties are found to be conservative due to the limited statistical precision of the \MGMCatNLO samples.
The hadronization and fragmentation modeling uncertainty is similarly estimated in both \ttbar and \Wt
by comparing the nominal \PowhegPythia sample with the same \PowhegHerwig sample.
The uncertainty due to the choice of parameters in the \PowhegPythia generator and P2012 underlying-event tune are derived by varying the parameters related to the
amount of initial- and final-state radiation, the factorization and renormalization scales,
and (for \ttbar only) the \pt of the first additional emission beyond the Born level~\cite{ATL-PHYS-PUB-2016-004}.
An uncertainty in the \singletop yield due to the destructive interference between the \ttbar and \Wt processes is estimated by
using inclusively generated $WWbb$ events in a comparison with the combined yield of \ttbar and \Wt samples, all generated at LO with \MGMCatNLO~2.5.5.

The theoretical uncertainties of the \Zjets, \diboson, and \ttV samples are estimated by varying event generator parameters
related to the factorization, renormalization, resummation, and CKKW matching scales.
The envelope of these variations is taken as the theoretical uncertainty in the predicted yield in each SR.
As the \diboson and \ttV samples are not normalized in the CRs, the uncertainty in the theoretical cross section is also included.
The uncertainty in the NLO cross section is taken to be 6\% for the \diboson process~\cite{ATL-PHYS-PUB-2016-002} and 13\% for the \ttV process~\cite{Alwall:2014hca}.
A 50\% uncertainty is applied to the small \Wjets yield in both SRs.

The \Stop signal model uncertainties are dominated by the cross-section uncertainty, derived from the envelope of
cross-section predictions from several distinct PDF sets and varying the factorization and renormalization scales,
as described in Ref.~\cite{Borschensky:2014cia}.
The uncertainty in the cross section varies from 13\% for the 600~\GeV mass value to 27\% for the 1600~\GeV mass value.
The electron efficiency uncertainties are between 3 and 4\% for the various \Stop masses when assuming $\BRe=\BRm=50\%$, and are between 5 and 8\% when assuming $\BRe=100\%$.
Similarly, the muon efficiency uncertainties are between 2 and 4\% when assuming $\BRe=\BRm=50\%$, and rise to 6\% when assuming $\BRm=100\%$.
The electron, muon, and jet energy scale and resolution uncertainties are generally below $1\%$ for the \Stop signal models,
 reaching 1\% for masses near the \mbl threshold of 800~\GeV for SR800 and 1100~\GeV for SR1100.
The $b$-tagging efficiency uncertainties are between 1 and 3\%, reaching the largest value for the 600~\GeV signal model.

\section{Results}
\label{sec:result}
The observed yields and fitted background predictions in SR800 and SR1100 are shown in \tabref{HistFitter_results_SR}.
One event is observed in SR1100 and two are observed in SR800, in agreement with the SM prediction.
The SR1100 event is included in SR800 by definition, and both events are found in the $\mu\mu$ channel.
The SR1100 event has a high \Ht due to a high-\pt muon with a large \pt uncertainty.
The observed and predicted $\Mbl^0$, \Ht, \MblAsym, \mll, and \Mblrej distributions in SR800 are shown in \figref{N1}.

\begin{table}[t]
\begin{center}
  \caption{
    The observed and total post-fit expected background yields in SR800 and SR1100.
    Both the MC background expectation before the fit and the background-only post-fit yields are shown, with each broken down
    into \singletop, \Zjet, \ttbar, \diboson, \ttV and \Wjet background processes.
    Model-independent upper limits are set at a 95\% CL on the visible number of expected ($S^{95}_\mathrm{exp}$) and observed ($S^{95}_\mathrm{obs}$) events
    and on the visible cross section ($\sigma_\mathrm{vis}$) of a generic BSM process.
    Results are shown in each flavor channel and inclusively.
    The background estimates and their uncertainties are derived from a background-only fit configuration.
    }
  \label{tab:HistFitter_results_SR}
\setlength{\tabcolsep}{0.0pc}
{\scriptsize
\begin{tabular*}{\textwidth}{@{\extracolsep{\fill}}lcccccccc}
\toprule
 & \multicolumn{4}{c}{SR800} & \multicolumn{4}{c}{SR1100} \\
           & inclusive    & $ee$   & $e\mu$   & $\mu\mu$
           & inclusive    & $ee$   & $e\mu$   & $\mu\mu$  \\[-0.05cm]
 \hline
 \hline
\rule{0pt}{1ex} Observed yield          & 2 & 0 & 0 & 2 & 1  & 0 & 0 & 1  \\
\rule{0pt}{2.5ex} Total post-fit bkg yield      & $5.2 \pm 1.4$  & $1.8 \pm 0.5$   & $2.1 \pm 0.8$   & $1.35 \pm 0.32$  &
                          $1.2_{-0.5}^{+0.6}$  & $0.51_{-0.20}^{+0.22}$ & $0.44_{-0.33}^{+0.39}$ & $0.22 \pm 0.13$ \\
 \hline
\rule{0pt}{1ex} Post-fit \singletop yield   & $2.0 \pm 1.3$           & $0.6 \pm 0.4$    & $1.1 \pm 0.7$     & $0.32 \pm 0.20$
                            & $0.32 \pm 0.29$         & $0.11 \pm 0.10$  & $0.21 \pm 0.19$   & --   \\
\rule{0pt}{1ex} Post-fit \Zjets yield       & $1.40 \pm 0.33$         & $0.80 \pm 0.24$  & $0.01 \pm 0.01$   & $0.59 \pm 0.14$
                            & $0.47 \pm 0.15$         & $0.28 \pm 0.10$  & --                & $0.19 \pm 0.11$   \\
\rule{0pt}{1ex} Post-fit \ttbar yield       & $1.0 \pm 0.5$           & $0.27 \pm 0.14$  & $0.54 \pm 0.25$   & $0.21 \pm 0.10$
                            & $0.21_{-0.21}^{+0.55}$  & $0.06_{-0.06}^{+0.16}$  & $0.13_{-0.13}^{+0.34}$   & $0.01_{-0.01}^{+0.03}$  \\
\rule{0pt}{1ex} Post-fit \diboson yield     & $0.64 \pm 0.23$         & $0.14 \pm 0.05$  & $0.31 \pm 0.12$   & $0.19 \pm 0.08$
                            & $0.13 \pm 0.05$         & $0.06 \pm 0.03$  & $0.07 \pm 0.03$   & $0.01 \pm 0.01$    \\
\rule{0pt}{1ex} Post-fit \ttV yield         & $0.12 \pm 0.03$         & $0.01 \pm 0.01$  & $0.07 \pm 0.02$   & $0.04 \pm 0.02$
                            & $0.03 \pm 0.01$         & --               & $0.01 \pm 0.01$   & $0.01 \pm 0.01$    \\
\rule{0pt}{1ex} Post-fit \Wjets yield       & $0.03 \pm 0.03$         & --               & $0.04 \pm 0.04$   & --
                            & $0.01_{-0.01}^{+0.02}$  & --               & $0.01_{-0.01}^{+0.02}$   & --    \\

 \hline
 \hline
\rule{0pt}{2.5ex} Total MC bkg yield    & $4.9 \pm 1.2$ & $1.7 \pm 0.4$   & $2.0 \pm 0.7$   & $1.23 \pm 0.28$
                & $1.1_{-0.5}^{+0.6}$ & $0.46_{-0.19}^{+0.21}$ & $0.43_{-0.33}^{+0.40}$ & $0.18 \pm 0.10$  \\
 \hline
\rule{0pt}{1ex} MC \singletop yield  &  $1.9 \pm 1.0$           & $0.57 \pm 0.34$         & $1.0 \pm 0.6$          & $0.29 \pm 0.17$
                     &  $0.29 \pm 0.25$         & $0.10 \pm 0.08$         & $0.19 \pm 0.17$        & --        \\
\rule{0pt}{1ex} MC \Zjets yield      &  $1.15 \pm 0.21$         & $0.65 \pm 0.17$         & $0.01 \pm 0.01$        & $0.48 \pm 0.09$
                     &  $0.38 \pm 0.10$         & $0.23 \pm 0.07$         & --                     & $0.15 \pm 0.09$        \\
\rule{0pt}{1ex} MC \ttbar yield      &  $1.1 \pm 0.5$           & $0.29 \pm 0.14$         & $0.57 \pm 0.26$        & $0.22 \pm 0.10$
                     &  $0.22_{-0.22}^{+0.57}$  & $0.07_{-0.07}^{+0.18}$  & $0.14_{-0.14}^{+0.36}$ & $0.01_{-0.01}^{+0.03}$ \\
\rule{0pt}{1ex} MC \diboson yield    &  $0.64 \pm 0.23$         & $0.14 \pm 0.05$         & $0.31 \pm 0.12$        & $0.19 \pm 0.08$
                     &  $0.13 \pm 0.05$         & $0.06 \pm 0.03$         & $0.07 \pm 0.03$        & $0.01 \pm 0.01$        \\
\rule{0pt}{1ex} MC \ttV yield        &  $0.12 \pm 0.03$         & $0.01 \pm 0.01$         & $0.07 \pm 0.02$        & $0.04 \pm 0.02$
                     &  $0.03 \pm 0.01$         & --                      & $0.01 \pm 0.01$        & $0.01 \pm 0.01$        \\
\rule{0pt}{1ex} MC \Wjets yield      &  $0.03 \pm 0.03$         & --                      & $0.04 \pm 0.04$        & --
                     &  $0.01_{-0.01}^{+0.02}$  & --                      & $0.01_{-0.01}^{+0.02}$ & --  \\
 \hline
 \hline
\rule{0pt}{2.5ex}  $S^{95}_\mathrm{exp}$ & ${ 6.4 }^{ +3.0 }_{ -1.9 }$
                    & $ { 4.1 }^{ +1.8 }_{ -1.1 }$
                    & $ { 4.0 }^{ +2.2 }_{ -0.9 }$
                    & $ { 3.9 }^{ +1.6 }_{ -0.7 }$
                    & $ { 3.9 }^{ +2.4 }_{ -0.5 }$
                    & $ { 3.0 }^{ +1.3 }_{ -0.0 }$
                    & $ { 3.0 }^{ +1.3 }_{ -0.0 }$
                    & $ { 3.1 }^{ +0.6 }_{ -0.1 }$ \\
\rule{0pt}{2.5ex}  $S^{95}_\mathrm{obs}$ & $4.0$ & $3.0$ & $3.0$ & $4.8$
                    & $3.9$ & $3.0$ & $3.1$ & $4.1$ \\
\rule{0pt}{2.5ex}  $\sigma_\mathrm{vis}$[fb]
                    & $0.11$ & $0.08$ & $0.08$ & $0.13$
                    & $0.11$ & $0.08$ & $0.08$ & $0.11$ \\
\bottomrule
\end{tabular*}
}
\end{center}
\end{table}

For each SR, model-independent upper limits are derived on the visible cross section of potential BSM processes at a 95\% confidence level (CL).
A likelihood fit is performed to the number of observed events in all three CRs and the target SR, and a generic BSM process
is assumed to contribute to the SR only.
No theoretical or systematic uncertainties are considered for the signal model except the luminosity uncertainty.
The observed ($S^{95}_\mathrm{obs}$) and expected ($S^{95}_\mathrm{exp}$) limits on the number of BSM events
are derived at 95\% CL in each flavor channel and inclusively, and are shown in the lower rows of \tabref{HistFitter_results_SR}.
Also shown are the observed limits on the visible cross section $\sigma_\mathrm{vis}$, defined as $S^{95}_\mathrm{obs}$ normalized to the integrated luminosity,
and representing the product of the production cross section, acceptance, and selection efficiency of a generic BSM signal.
Limits on $\sigma_\mathrm{vis}$ are set between 0.08 and 0.13 fb, with the weaker limit set in the $\mu\mu$ channel due to the two observed events.

Exclusion limits are derived at 95\% CL for the \StopSym signal samples.
Limits are obtained through a profile log-likelihood ratio test using the \CLs prescription~\cite{Read:2002hq},
following the simultaneous fit to the CRs and a target SR~\cite{Baak:2014wma}.
The signal contributions in both the SR and CRs are accounted for in the fit, although they are negligible in the latter.
Exclusion fits are performed separately for various branching ratio assumptions, sampling
values of \BRe, \BRm, and \BRt whose sum is unity in steps of 5\%, and reweighting events in the signal samples according to the generated decays.
For both SR800 and SR1100, limits are derived in the $ee$, $e\mu$, $\mu\mu$, and inclusive channels.
Observed limits are reported for the SR and channel combination with the lowest expected \CLs value, and therefore best expected sensitivity,
at a given mass value and branching ratio.
The inclusive channel typically has the stronger expected sensitivity when \BRe and \BRm are both above 15\%,
while the $ee$ ($\mu\mu$) channel is more sensitive when \BRm (\BRe) is below 15\%.
The inclusive channel is always more sensitive than the $e\mu$ channel because a substantial fraction of signal events have two leptons of the same flavor, regardless of individual branching ratios.

The expected and observed exclusion contours for the branching ratios
are shown in \figref{limit_contours} for each simulated \StopSym mass.
The limits are strongest at low values of \BRt, where the expected number of
events with electrons or muons in the final state is largest.
Expected limits are slightly stronger for increasing \BRe, reflecting a higher
trigger efficiency for electrons than for muons.
Stops with \BRt up to 80\% or more are excluded for masses between 600 and 1000~\GeV, while
those with larger \BRe or \BRm may be excluded up to 1500~\GeV.
Observed limits are stronger than expected for \StopSym masses of 1100~\GeV or below,
reflecting the lower-than-expected event yield in SR800 in the $ee$ channel and inclusively.
Exclusion contours reflecting the highest \StopSym mass excluded at a 95\% CL for a given
point in the branching ratio plane are shown in \figref{mass_limit_obs}.

\begin{figure}[!h]
  \centering
  \includegraphics[width=\textwidth]{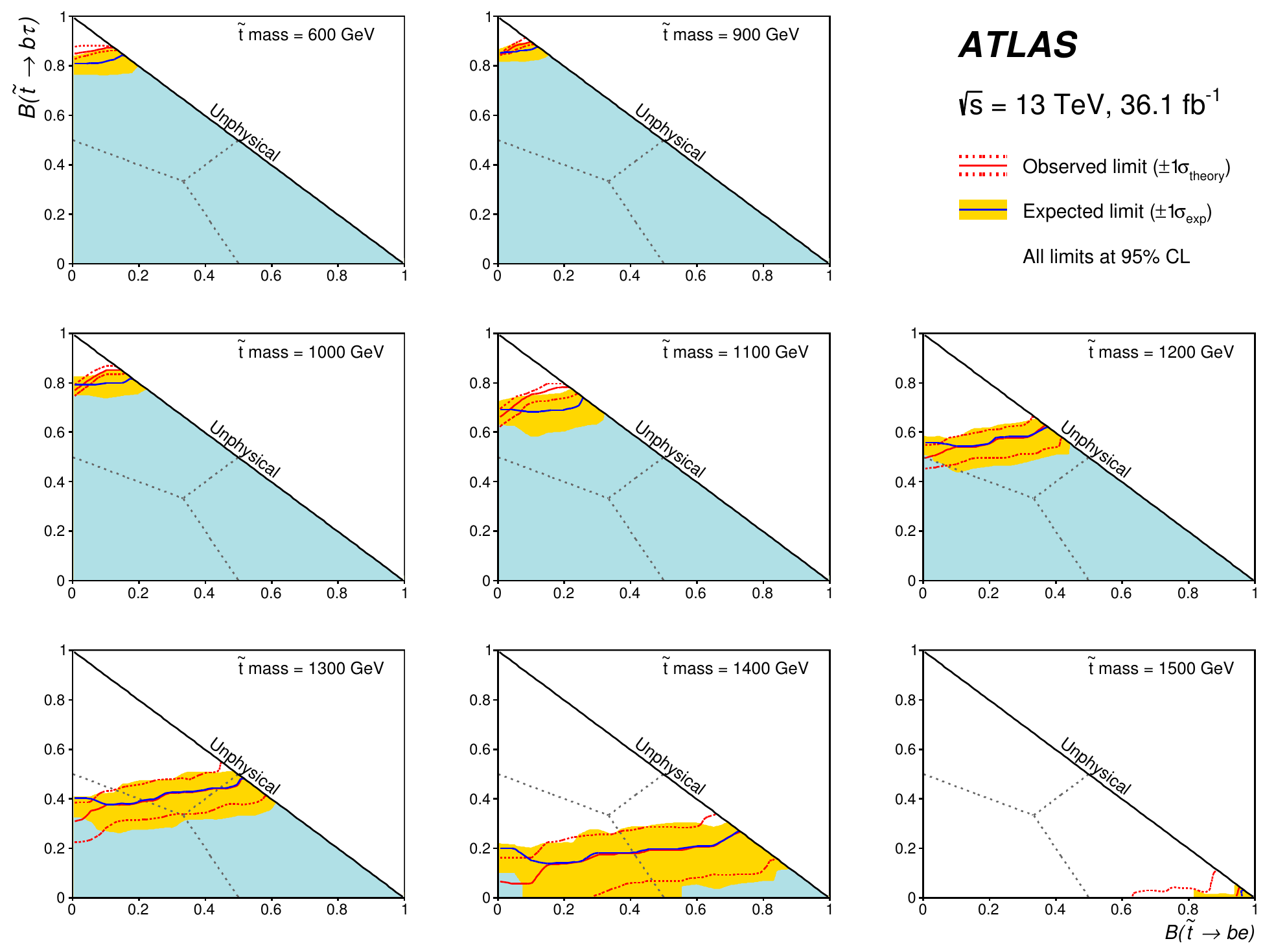}
  \caption{Expected (dashed blue) and observed (solid red) limit curves as a function of \StopSym branching ratios
    for various mass values between 600 and 1500~\GeV.
    The sum of \BRe, \BRm, and \BRt is assumed to be unity everywhere, and points of equality are marked by a dotted gray line.
    The yellow band reflects the $\pm 1\sigma$ uncertainty of the expected limit due to theoretical, experimental, and MC statistical uncertainties.
    The shaded blue area represents the branching ratios that are expected to be excluded beyond $1\sigma$.
    The dotted red lines correspond to the $\pm 1\sigma$ cross section uncertainty of the observed limit
    derived by varying the signal cross section by the theoretical uncertainties.
  }
  \label{fig:limit_contours}
\end{figure}

\begin{figure}[!h]
  \centering
  \includegraphics[width=0.6\textwidth]{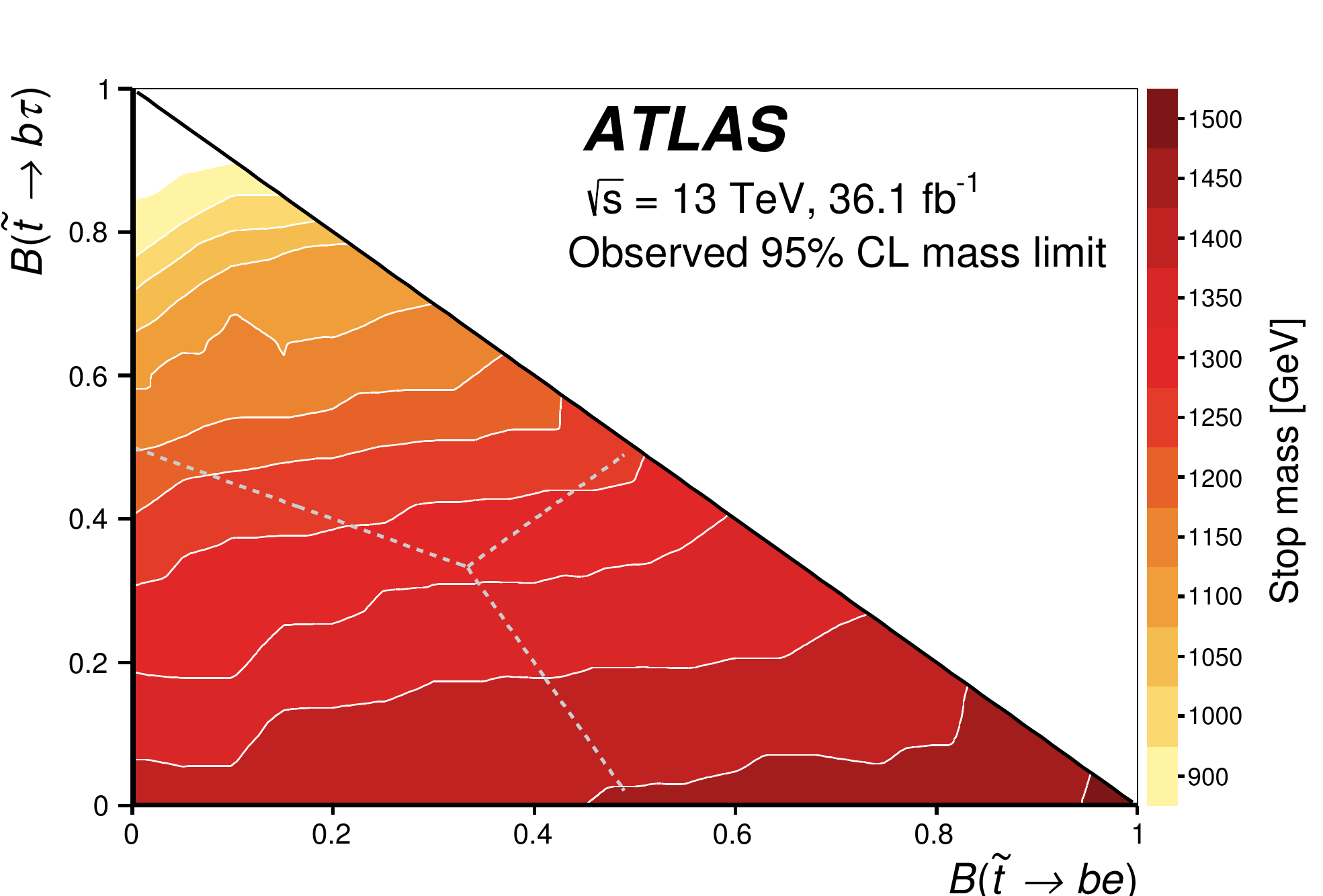}
  \caption{The observed lower limits on the \StopSym mass at 95\% CL as a function of \StopSym branching ratios.
    The sum of \BRe, \BRm, and \BRt is assumed to be unity everywhere, and points of equality are marked by a dotted gray line.
    The limits are obtained using the nominal \StopSym cross-section predictions.
    As the branching ratio \BRt increases, the
    expected number of events with electrons or muons in the final state
    decreases, reducing the mass reach of the exclusion.
  }
  \label{fig:mass_limit_obs}
\end{figure}

\FloatBarrier

\section{Conclusions}
\label{sec:conclusion}

This paper presents the first ATLAS results on the search for the pair production of \Stops, each decaying
via an $R$-parity-violating coupling to a $b$-quark and a lepton.
The final state requires two jets, at least one of which is $b$-tagged, and two light, oppositely-charged leptons (electron or muon).
The search uses 36.1~\ifb\ of $\sqrt{s}=13~\TeV$
proton--proton collision data collected with the ATLAS detector at the LHC in 2015 and 2016.
No significant excess of events over the Standard Model prediction is observed,
and limits are set on the \StopSym mass at a 95\% confidence level.
These results significantly extend the lower-mass exclusion limits on the \BL \Stop model from reinterpretations of Run 1 leptoquark searches.
Model-independent upper limits are set on the cross section of potential BSM processes in the $ee$, $e\mu$, and $\mu\mu$ channels and inclusively.
A scan of various \StopSym branching ratios is performed to set branching-ratio-dependent limits on decays to $be$, $b\mu$, and $b\tau$ for various \StopSym mass models.
Limits are set on \StopSym masses between 600~\GeV for large $b\tau$ decay branching ratios and 1500~\GeV for a $be$ branching ratio of 100\%.

\section*{Acknowledgments}
We thank CERN for the very successful operation of the LHC, as well as the
support staff from our institutions without whom ATLAS could not be
operated efficiently.

We acknowledge the support of ANPCyT, Argentina; YerPhI, Armenia; ARC, Australia; BMWFW and FWF, Austria; ANAS, Azerbaijan; SSTC, Belarus; CNPq and FAPESP, Brazil; NSERC, NRC and CFI, Canada; CERN; CONICYT, Chile; CAS, MOST and NSFC, China; COLCIENCIAS, Colombia; MSMT CR, MPO CR and VSC CR, Czech Republic; DNRF and DNSRC, Denmark; IN2P3-CNRS, CEA-DRF/IRFU, France; SRNSF, Georgia; BMBF, HGF, and MPG, Germany; GSRT, Greece; RGC, Hong Kong SAR, China; ISF, I-CORE and Benoziyo Center, Israel; INFN, Italy; MEXT and JSPS, Japan; CNRST, Morocco; NWO, Netherlands; RCN, Norway; MNiSW and NCN, Poland; FCT, Portugal; MNE/IFA, Romania; MES of Russia and NRC KI, Russian Federation; JINR; MESTD, Serbia; MSSR, Slovakia; ARRS and MIZ\v{S}, Slovenia; DST/NRF, South Africa; MINECO, Spain; SRC and Wallenberg Foundation, Sweden; SERI, SNSF and Cantons of Bern and Geneva, Switzerland; MOST, Taiwan; TAEK, Turkey; STFC, United Kingdom; DOE and NSF, United States of America. In addition, individual groups and members have received support from BCKDF, the Canada Council, CANARIE, CRC, Compute Canada, FQRNT, and the Ontario Innovation Trust, Canada; EPLANET, ERC, ERDF, FP7, Horizon 2020 and Marie Sk{\l}odowska-Curie Actions, European Union; Investissements d'Avenir Labex and Idex, ANR, R{\'e}gion Auvergne and Fondation Partager le Savoir, France; DFG and AvH Foundation, Germany; Herakleitos, Thales and Aristeia programmes co-financed by EU-ESF and the Greek NSRF; BSF, GIF and Minerva, Israel; BRF, Norway; CERCA Programme Generalitat de Catalunya, Generalitat Valenciana, Spain; the Royal Society and Leverhulme Trust, United Kingdom.

The crucial computing support from all WLCG partners is acknowledged gratefully, in particular from CERN, the ATLAS Tier-1 facilities at TRIUMF (Canada), NDGF (Denmark, Norway, Sweden), CC-IN2P3 (France), KIT/GridKA (Germany), INFN-CNAF (Italy), NL-T1 (Netherlands), PIC (Spain), ASGC (Taiwan), RAL (UK) and BNL (USA), the Tier-2 facilities worldwide and large non-WLCG resource providers. Major contributors of computing resources are listed in Ref.~\cite{ATL-GEN-PUB-2016-002}.

\printbibliography
\newpage

\begin{flushleft}
{\Large The ATLAS Collaboration}

\bigskip

M.~Aaboud$^\textrm{\scriptsize 137d}$,
G.~Aad$^\textrm{\scriptsize 88}$,
B.~Abbott$^\textrm{\scriptsize 115}$,
O.~Abdinov$^\textrm{\scriptsize 12}$$^{,*}$,
B.~Abeloos$^\textrm{\scriptsize 119}$,
S.H.~Abidi$^\textrm{\scriptsize 161}$,
O.S.~AbouZeid$^\textrm{\scriptsize 139}$,
N.L.~Abraham$^\textrm{\scriptsize 151}$,
H.~Abramowicz$^\textrm{\scriptsize 155}$,
H.~Abreu$^\textrm{\scriptsize 154}$
% ...
\bigskip
\\
% ...
$^{12}$ Institute of Physics, Azerbaijan Academy of Sciences, Baku, Azerbaijan\\
% ...
$^{88}$ CPPM, Aix-Marseille Universit{\'e} and CNRS/IN2P3, Marseille, France\\
% ...
$^{115}$ Homer L. Dodge Department of Physics and Astronomy, University of Oklahoma, Norman OK, United States of America\\
% ...
$^{119}$ LAL, Univ. Paris-Sud, CNRS/IN2P3, Universit{\'e} Paris-Saclay, Orsay, France\\
% ...
$^{137}$ $^{(a)}$ Facult{\'e} des Sciences Ain Chock, R{\'e}seau Universitaire de Physique des Hautes Energies - Universit{\'e} Hassan II, Casablanca; $^{(b)}$ Centre National de l'Energie des Sciences Techniques Nucleaires, Rabat; $^{(c)}$ Facult{\'e} des Sciences Semlalia, Universit{\'e} Cadi Ayyad, LPHEA-Marrakech; $^{(d)}$ Facult{\'e} des Sciences, Universit{\'e} Mohamed Premier and LPTPM, Oujda; $^{(e)}$ Facult{\'e} des sciences, Universit{\'e} Mohammed V, Rabat, Morocco\\
% ...
$^{139}$ Santa Cruz Institute for Particle Physics, University of California Santa Cruz, Santa Cruz CA, United States of America\\
% ...
$^{151}$ Department of Physics and Astronomy, University of Sussex, Brighton, United Kingdom\\
% ...
$^{154}$ Department of Physics, Technion: Israel Institute of Technology, Haifa, Israel\\
$^{155}$ Raymond and Beverly Sackler School of Physics and Astronomy, Tel Aviv University, Tel Aviv, Israel\\
% ...
$^{161}$ Department of Physics, University of Toronto, Toronto ON, Canada\\
% ...
$^{a}$ Also at Department of Physics, King's College London, London, United Kingdom\\
$^{b}$ Also at Institute of Physics, Azerbaijan Academy of Sciences, Baku, Azerbaijan\\
$^{c}$ Also at Novosibirsk State University, Novosibirsk, Russia\\
$^{d}$ Also at TRIUMF, Vancouver BC, Canada\\
$^{e}$ Also at Department of Physics {\&} Astronomy, University of Louisville, Louisville, KY, United States of America\\
% ...
$^{*}$ Deceased
\end{flushleft}


\end{document}